\documentclass[12pt]{article}

\usepackage{amsmath}         % \
\usepackage{amssymb}         %  | AMS Packages for math
\usepackage{amsfonts}        % /
\usepackage{pifont}          % from Brando
\usepackage{graphicx}        % Graphics
\usepackage{bbm}             % for \mathbbm{1} (unit matrix)
\usepackage{epstopdf}

\usepackage{color}         % for \color{...}, colored text
\usepackage{slashed}       % for \slashed{k}
\usepackage{framed}        % for remarks
\usepackage{mathrsfs}      % for Weinberg-esque letters
\usepackage{ragged2e}      % for ragged right (for institutions)
\usepackage{lipsum}        % for block of text (formatting test)
\usepackage{paralist}      % for compactitem
\usepackage{multirow}      % for multiple row elements in a table
\usepackage{cite}          % cancels collref, but groups citations

\usepackage{caption}
\usepackage{subcaption}

\newcommand{\be}{\begin{eqnarray}}
\newcommand{\ee}{\end{eqnarray}}

\usepackage[margin=2cm]{geometry} % for reasonable margins

\usepackage{scalefnt}			    % for \smaller
\graphicspath{{figures/}}	    % separate folder for figures

\numberwithin{equation}{section}

\newenvironment{institutions}[1][2em]
  {\begin{list}{}{\setlength\leftmargin{#1}\setlength\rightmargin{#1}}\item[]\RaggedRight}
  {\end{list}}

\usepackage[
	colorlinks=true,
	citecolor=black,
	linkcolor=black,
	urlcolor=blue,
	hypertexnames=false]{hyperref}

\begin{document}

\thispagestyle{empty}
\begin{center}

	{	% TITLE GOES HERE 	
		\huge \bf 
		WIMP Dark Matter through the Dilaton Portal
	}

	\vskip .7cm
	
	\renewcommand*{\thefootnote}{\fnsymbol{footnote}}
	% Sets footnotes to a sequence of symbols, for e-mails

	{	% AUTHORS GO HERE
		\bf
		Kfir Blum$^{a}$,
		Mathieu Cliche$^{b}$,
		Csaba Cs\'aki$^{b}$,
		and Seung J. Lee$^{c,d}$
	}  

	\renewcommand{\thefootnote}{\arabic{footnote}}
	\setcounter{footnote}{0}
	% Returns footnote to numbers and restarts counter

	\vspace{.2cm}

\begin{institutions}[1.5cm]
\footnotesize

    % UNIVERSITY IDENTIFICATIONS HERE
    
	$^{a}$ {\it Institute for Advanced Study, Princeton,  \textsc{nj} 08540,  \textsc{usa}} \\

	\vspace*{0.05cm}

	$^{b}$ {\it Department of Physics, \textsc{lepp}, Cornell University, Ithaca, \textsc{ny} 14853, \textsc{usa}} \\

\vspace*{0.05cm}

 $^{c}$ {\it Department of Physics, Korea Advanced Institute of Science and Technology, 335 Gwahak-ro, Yuseong-gu, Daejeon 305-701, \textsc{Korea}}\\

	\vspace*{0.05cm}

$^{d}$ {\it School of Physics, Korea Institute for Advanced Study, Seoul 130-722, \textsc{Korea}}

\end{institutions}
	
\end{center}

%%%%%%%%%%%%%%%%%%%%%%%%%%%%%%%%%%%%%%%%%%%%%%%%%%%%%%%%%%%%
%One of these parameters is further fixed by requiring that thermal freeze-out results in the observed dark matter relic abundance. 

\begin{abstract}
\noindent 
We study a model in which dark matter couples to the Standard Model through a dilaton of a sector with spontaneously broken approximate scale invariance. Scale invariance fixes the dilaton couplings to the Standard Model and dark matter fields, leaving three main free parameters: the symmetry breaking scale $f$, the dilaton mass $m_{\sigma}$, and the dark matter mass $m_{\chi}$.  
We analyze the experimental constraints on the parameter space from collider, direct and indirect detection experiments including the effect of Sommerfeld enhancement, and show that dilaton exchange provides a consistent, calculable framework for cold dark matter with $f,\,m_\sigma,\,m_\chi$ of roughly similar magnitude and in the range $\sim 1-10$ TeV. Direct and indirect detection experiments, notably future ground-based gamma ray and space-based  cosmic ray measurements, can probe the model all the way to dark matter mass in the multi-TeV regime.
%We show that WIMP dark matter that couples to the Standard Model through dilaton exchange can be explored in future direct and indirect detection experiments up to multi-TeV dark matter mass. %, including  sensitive to dark matter mass at and above the TeV scale, as well as in future indirect experiments such as high energy ground-based gamma ray measurements and antiproton flux through the improved high energy cosmic ray measurements.

\end{abstract}
%\tableofcontents
%%%%%%%%%%%%%%%%%%%%%%%%%%%%%%%%%%%%%%%%%%%%%%%%%%%%%%%%%%%%

\section{Introduction}

Embedding the Standard Model (SM) partially or completely in a composite sector can solve the hierarchy problem, by making the Higgs boson composite. Often such a composite sector arises as the low-energy limit of an approximately scale invariant theory, where scale invariance is broken somewhere above the weak scale. If the breaking of scale invariance is spontaneous, then it is accompanied by a light dilaton $\sigma$ that couples to the fields in the composite sector through the trace of the energy-momentum tensor~\cite{CGRT, GRW, CGK, CHL, Goldberger:2008zz, Eshel:2011wz} 
\begin{equation}\label{eq:TrT}
-\frac{\sigma}{f} {\rm Tr} T.
\end{equation}
For massive particles, the coupling to $\sigma$ is proportional to the particle masses, with the suppression scale $f$ corresponding to the breaking of scale invariance. 

The canonical dilaton Lagrangian (\ref{eq:TrT}) offers an economical way to couple the SM to new fields that could be singlets under the SM gauge symmetries and thus form an otherwise dark sector. 
In this paper we study the possibility that dark matter (DM) belongs to such dark sector and couples to the SM through Eq.~(\ref{eq:TrT}). 
In the minimal set up that we explore here, three parameters determine the dynamics of thermal freeze-out in the early Universe: the breaking scale $f$, the dilaton mass $m_\sigma$, and the dark matter mass $m_\chi$. Fixing one of these parameters such that the observed dark matter relic abundance is reproduced leaves a rather predictive framework. 
We show that a large parametric region exists where the solution is perturbative and produces cold, weakly interacting massive particle dark matter (WIMP), with $f,\,m_\sigma,\,m_\chi$ of roughly similar magnitude and in the range $\sim 1-10$~TeV. 

Null results from dark matter direct detection experiments like LUX \cite{Akerib:2013tjd}, XENON100 \cite{Aprile:2012nq} and CDMS\cite{Agnese:2013rvf} put considerable pressure on WIMP models where DM couples to the SM through exchange of SM particles. The annihilation cross section $\sigma_{\rm ann}\sim10^{-36}$~cm$^2$, required for WIMP relic abundance consistent with observations, is some ten orders of magnitude larger than the WIMP-nucleon elastic scattering cross section now probed by the direct detection experiments. This excludes $Z$ boson exchange in all but fine-tuned corners of the parameter space, and requires some tuning for Higgs mediation as well. In contrast, the dilaton portal we analyze here quite generically evades the direct detection constraints in the bulk of the relevant parameter space, as the DM coupling to the SM resembles the case of Higgs exchange but with extra suppression of order $(v/f)^2\,(m_h/m_\sigma)^4$ with $v=\langle H\rangle=246$~GeV and $m_\sigma$ and the scale $f$ automatically lying in the TeV ballpark to provide the correct relic abundance.
 
The idea that dark matter could couple to the SM via dilaton exchange was analyzed previously in Ref.~\cite{Bai:2009ms} (where the dilaton was taken to be massless) and in Ref.~\cite{Agashe:2009ja} (for some specific warped extra dimensional models where the role of the dilaton was played by the radion). Our work generalizes the results of Ref.~\cite{Agashe:2009ja} and extends the analysis of~\cite{Bai:2009ms} by adding the dilaton mass as a free parameter. This allows a more complete exploration of the parameter space and reveals effects such as Sommerfeld-enhanced annihilation. We also incorporate the most recent experimental bounds from direct and indirect detection as well as collider experiments.

The outline of this paper is as follows. In Sec.~\ref{sec:mod} we summarize the basic properties of the dilaton. We present its couplings, fixed mainly by the scale $f$ with a few additional parameters characterizing the embedding of the SM matter into the composite sector, comment on expected NDA bounds on the dilaton mass, and present two benchmark models to be studied in the paper. Sec.~\ref{sec:ra} contains a calculation of the DM annihliation cross section due to dilaton exchange. After deriving a unitarity bound on the DM mass, we present the parameter space of the theory where the observed relic abundance is reproduced. In Sec.~\ref{sec:dd} we compare the DM-nucleon scattering cross sections to the experimental bounds from the latest round of direct detection measurements, finding that large regions of the parameter space are compatible with the bounds. In Sec.~\ref{sec:id} we consider constraints from indirect detection of gamma rays and cosmic ray antiprotons. We conclude in Sec.~\ref{sec:conc}. App. A summarizes collider bounds on the dilaton, considering LEP, Tevatron and the LHC. App. B contains cross-section formulae for the sub-leading annihilation channels that we omit in the body of the text for clarity.

%%%%%%%%%%%%%%%%%%%%%%%%%%%%%%%%%%%%%%%%%
\section{The Dilaton Mediated Dark Matter Model} \label{sec:mod}

We start by considering the effective theory describing an approximately scale invariant sector with scale invariance spontaneously broken at the scale $f$. The Goldstone boson corresponding to this breaking, called the dilaton $\sigma(x)$, can be parametrized via a spurion field $\Phi(x)$ as \cite{Bellazzini:2012vz}
\begin{eqnarray}
\Phi(x) \equiv f e^{\sigma(x) /f}
\end{eqnarray}
such that under a scale transformation $x\to x e^\lambda$ we have $\Phi(x)\rightarrow e^{\lambda}\Phi(e^{\lambda}x)$ and $\langle\Phi\rangle=f$.  To obtain a canonically normalized dilaton kinetic term it is convenient to do a field redefinition such that $\Phi(x)=\sigma(x)+f$ \cite{Goldberger:2008zz} where $\sigma $ is now the canonically normalized dilaton field.  Using a spurion analysis one can then find the low energy theory below the cutoff scale $4\pi f$ by inserting powers of $\Phi/f$ in the SM Lagrangian to make it scale invariant.  After electroweak symmetry breaking one finds the following effective action describing the interactions of the canonically normalized dilaton with the SM fields~\cite{CHL,Goldberger:2008zz,Bellazzini:2012vz} 
\begin{eqnarray}
\mathcal{L}_{\sigma}\label{eq:Leff}&=&\frac{1}{2}\partial_{\mu}\sigma\partial^{\mu}\sigma-\frac{1}{2}m_{\sigma}^2\sigma^2-\frac{5}{6}\frac{m_{\sigma}^2}{f}\sigma^3-\frac{11}{24} \frac{m_\sigma^2}{f^2} \sigma^4+\ldots -\left(\frac{\sigma}{f}\right)\left[\sum_{\psi} (1+\gamma_\psi )m_{\psi}\bar{\psi}\psi\right]\nonumber +\\  &+&\left(\frac{2\sigma}{f}+\frac{\sigma^2}{f^2}\right)\left[m_{W}^2W^{+\mu}W^{-}_{\mu}+\frac{1}{2}m_{Z}^2Z^{\mu}Z_{\mu}-\frac{1}{2}m_h^2h^2\right]+
\frac{\alpha_{\text{EM}}}{8\pi f}c_{\text{EM}}\sigma F_{\mu\nu}F^{\mu\nu}+\nonumber \\ &+&\frac{\alpha_{\text{s}}}{8\pi f}c_{\text{G}}\sigma G_{a\mu\nu}G^{a\mu\nu}.
\end{eqnarray}
The sum on $\psi$ runs over the SM fermions, which are assumed to be partially composite with light fermions being mainly elementary and the top quark mainly composite. $\gamma_\psi$ corresponds to the anomalous dimension of fermionic operators responsible for generating the SM fermion masses after mixing between the elementary and composite sectors. For composite fermions the anomalous dimension is expected to be small $\gamma_\psi \simeq 0$, while for light fermions the anomalous dimension may be sizable.

Naive dimensional analysis (NDA) limits the plausible size of the dilaton mass. For example, considering the dilaton self-energy loop from the trilinear coupling in Eq.~(\ref{eq:Leff}) we find that 
\begin{equation}
m_\sigma \leq 4 \pi f
\label{eq:NDA}
\end{equation}
to ensure that the one-loop correction of the dilaton mass remains below the tree-level mass, and that the couplings of the dilaton to matter remain under control~\cite{Chacko:2012sy}. This is just the reflection of the fact that this theory has an intrinsic cutoff of order $\Lambda \sim 4 \pi f$, and we should treat it as an effective theory valid below that scale. Note also that it is difficult to make the dilaton much lighter than the cutoff scale. In generic models there is a tuning of order $\frac{m_\sigma} {\Lambda}$ necessary to lower the dilaton mass~\cite{Bellazzini:2012vz,Chacko:2012sy,Chacko}, though special constructions can potentially alleviate this tuning~\cite{BCHST,Rattazzi}. We will require that the dilaton is not lighter than $f/10$. 

A few additional comments are in order about Eq.~(\ref{eq:Leff}).  The cubic and quartic dilaton self interaction arise from expanding the effective dilaton potential which includes a scale invariant term, $\Phi^4$, and small explicit sources of scale symmetry breaking such as $\Phi^{4-\epsilon}$.   Requiring that $\langle\Phi\rangle=f$ and that $\frac{d^2V(\Phi)}{d\Phi^2}=m_{\sigma}^2$ fixes the parameters of the dilaton potential. The leading expression for the cubic self-coupling of the dilaton is $5/6$ in the $\epsilon \rightarrow 0$ limit and the quartic is $11/24$.  Away from the $\epsilon \to 0$ limit, the cubic coupling can lie anywhere in the interval $[2/6,5/6]$ \cite{Goldberger:2008zz}. For simplicity, throughout this paper  we have used the limiting value $5/6$ for the cubic,  though we have verified that this does not influence our results significantly.  The coupling of the dilaton to massless gauge bosons arises from two sources; just like for the SM Higgs, the dilaton receives a contribution from top quark and $W$ boson loops, but in addition there is a direct contribution from the trace anomaly. The trace anomaly is proportional to the $\beta$-functions: the actual contribution will be the difference between the $\beta$-function above and below the symmetry breaking scale.   Thus this contribution depends on the details of what fraction of the composite sector is actually charged under the unbroken SM gauge symmetries, and what fraction of the SM fields are composites.   
For example, the coupling to gluons $c_G$ receives a contribution from the trace anomaly and from a top loop and is given by
\begin{equation}
c_G=b^{(3)}_{\text{IR}}-b^{(3)}_{\text{UV}}+\frac{1}{2}F_{1/2}(x_t)
\end{equation}
where $b^{(3)}_{UV,IR}$ are the QCD $\beta$-function coefficients above and below the scale $f$. This is a free parameter of the theory, which gives a measure of the QCD charges of the scale invariant sector. The function $F_{1/2}$ is the usual triangle diagram contribution of a fermion given by 
\begin{eqnarray}
F_{1/2}(x)&=& 2x\left[1+(1-x)f(x)\right]\\
f(x) &=& \left\{
     \begin{array}{lr}         \left[\sin^{-1}\left(1/\sqrt{x}\right)\right]^2 &  x\geq 1\\
 -\frac{1}{4}\left[\log\left(\frac{1+\sqrt{x-1}}{1-\sqrt{x-1}}\right)-i\pi\right]^2 &  x<1 \end{array}
   \right.
\end{eqnarray}
where $x_t=4m_t^2/m_{\sigma}^2$ \cite{CHL,Bellazzini:2012vz}.  A similar expression applies to the coupling to photons. 

Some of the results in the following sections (in particular the direct detection and collider signals)  depend on the parameters $c_G$ and $c_{EM}$.  To this end we define two benchmark model examples which we will study in detail. 
\begin{description}
  \item[Model A:] {This is the well-studied case proposed in~\cite{Goldberger:2008zz} where the entire SM is composite, corresponding to $b_{UV}=0, b_{IR}=b_{SM}$, giving rise to the parameters $b^3_{UV}-b^3_{IR}=-7, \  b^{EM}_{UV}-b^{EM}_{IR}=11/3$. Note that for a light dilaton these $b$'s depend somewhat on the dilaton mass: for example $b_{UV}^3-b_{IR}^3=-11+2n/3$, with $n$ denoting the number of quarks whose mass is smaller than $m_{\sigma}/2$.}
  \item[Model B:] {This is a limit of the well-motivated case when only the right-handed top and the Goldstone bosons needed for electroweak symmetry breaking are composites, while we minimize the $\beta$-functions of the UV to be as small as possible, resulting in
$b^3_{UV}=b^{EM}_{UV}=0, \ b^3_{IR}=-1/3, \ b^{EM}_{IR}=-11/9$. Note however that $b_{UV}$ is in fact a free parameter depending on the actual UV theory, and its value here has been chosen only for illustration.}
\end{description}

The final ingredient of the model is $\chi$, the dark matter particle, which can be spin 0, 1/2 or 1.  We assume that $\chi$ is a composite of the conformal sector, and does not have any direct coupling to the standard model fields which are mainly elementary.  The couplings of $\chi$ to the dilaton are fixed by a spurion analysis and follow the rules of couplings of generic massive composites\cite{Bai:2009ms}:
\begin{eqnarray}\label{eq:DMsig}
\mathcal{L}_{\text{DM}} \supset \left\{
     \begin{array}{lr}             -\left(1+\frac{2\sigma}{f}+\frac{\sigma^2}{f^2}\right)\frac{1}{2}m^2_{\chi}\chi^2 &  \text{Scalar}\\
  -\left(1+\frac{\sigma}{f}\right)m_{\chi}\bar{\chi}\chi &  \text{Fermion}\\
\left(1+\frac{2\sigma}{f}+\frac{\sigma^2}{f^2}\right)\frac{1}{2}m^2_{\chi}\chi_{\mu}\chi^{\mu} &  \text{Gauge boson.}
     \end{array}
   \right.
\end{eqnarray}
For simplicity we assume that a $\mathcal{Z}_2$ symmetry renders $\chi$ to be a stable particle.  For the fermionic case, we assume that $\chi$ is a Dirac fermion.

%%%%%%%%%%%%%%%%%%%%%%%%%%%%%%%%%%%%%%%%
\section{Relic Abundance} \label{sec:ra}

In this section we present the calculation of the relic abundance of the dark matter field $\chi$, where annihilations into SM states are assumed to proceed via dilaton exchange, and exhibit the relevant parameter space of the theory.  As usual, for small relative velocities $v$ the velocity-weighted annihilation cross section can be expanded  as $\sigma v = a + bv^2$. At the freeze-out temperature $T_F$  we have $\langle v^2\rangle = 6/x_F$ where $x_F=m_{\chi}/T_F$. The value of $x_F$ can then be determined by solving the Boltzmann equation in an expanding Universe:   
\begin{eqnarray}
x_F = \ln\left(\frac{5}{4}\sqrt{\frac{45}{8}}\frac{g}{2\pi^3}\frac{M_{\text{Pl}}m_{\chi}(a+6b/x_F)}{\sqrt{g^{*}}\sqrt{x_F}}\right),\label{eq:xF}
\end{eqnarray}
where $g$ is the number of degrees of freedom of the dark matter particle and $g^{*}$ is the effective number of relativistic degrees of freedom in thermal equilibrium during dark matter freeze-out. Once $x_F$ is determined  the dark matter relic abundance is given by
\begin{eqnarray}
\Omega_{\chi}h^2 \approx \frac{1.07\times 10^9}{\text{GeV}M_{\text{Pl}}\sqrt{g^{*}}}\frac{x_F}{a+3(b-a/4)/x_F}.
\end{eqnarray}

As we show below, the dark matter annihilation cross section (and thus the parameters $a, b$) in the model considered here is calculated in terms of  $m_{\chi}, m_{\sigma}$ and $f$.   Requiring that the observed relic abundance   $\Omega_{\chi}h^2 = 0.1199 \pm 0.0027$~\cite{Ade:2013zuv} is reproduced will thus impose one non-trivial relation and reduce the parameter space of the model. Next we map out this relation in detail, obtaining the reduced parameter space of the theory to be tested by direct and indirect detection experiments as well as collider searches. 

\subsection{Annihilation cross sections}

%%%%%%%%%%%%%%
\begin{figure}
\begin{center}
\includegraphics[width=.25\textwidth]{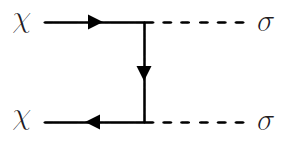}\;
\includegraphics[width=.25\textwidth]{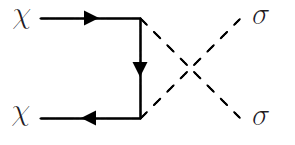}\;
\includegraphics[width=.25\textwidth]{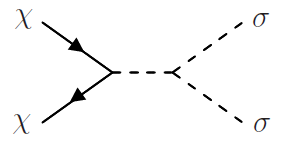}\;
\includegraphics[width=.25\textwidth]{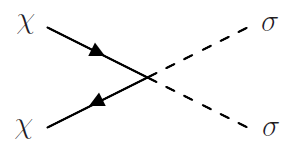}\;
\includegraphics[width=.25\textwidth]{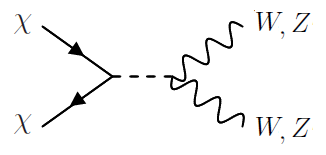}
\end{center}
\caption{Leading annihilation diagrams of dark matter in the regime $m_{\chi}\gg m_t$.  For fermionic dark matter there is no direct annihilation to dilatons.
}
\label{fig:diagrams}
\end{figure}
%%%%%%%%%%%%%%

The dominant dark matter annihilation channels for $m_\chi \gg m_t$ are $\chi\chi \to \sigma\sigma , WW, ZZ$, shown in  Fig. \ref{fig:diagrams}. The dominant channels contain factors of $m_{\chi}/f$, to be compared with all other sub-leading channels (for example s-channel dilaton exchange with quark or higgs final states) that contain factors of  $m_q/f$ or $m_h/f$ instead and are thus suppressed by relative powers of $m_{q,h}/m_\chi$. 
%For the case of dilaton final states the couplings themselves are directly proportional to $m_\chi /f$, while for the annihilation to $WW,ZZ$ the $m_{\chi}/f$ factors arise from the enhancement of the couplings of the longitudinal modes.  
Below we present analytical expressions for the dominant channels in the limit $m_{\chi}\gg m_t$, for the cases of scalar, fermion and vector dark matter. Formulae for the sub-leading annihilation channels can be found in Appendix \ref{app:A}. For numerical results all of the allowed annihilation channels are included.  

\subsubsection*{Scalar dark matter}

Scalar dark matter annihilation is dominated by s-wave processes. The approximate expressions of the cross sections are
\begin{eqnarray}
\sigma v(\chi\chi\rightarrow WW) &\simeq& \frac{m_{\chi}m_W^4\sqrt{m_{\chi}^2-m_{W}^2}\left(2+\frac{(2m_{\chi}^2-m_W^2)^2}{m_W^4}\right)}{2\pi f^4|4m_{\chi}^2-m_{\sigma}^2-i\text{Im}\left(\Pi(4m_{\chi}^2)\right)|^2}\label{eq:wwann},\\
\sigma v(\chi\chi\rightarrow \sigma\sigma) &\simeq& \frac{m_{\chi}\sqrt{m_{\chi}^2-m_{\sigma}^2}|2(2m_{\chi}^2-m_{\sigma}^2)^2+2m_\sigma^4-i\text{Im}\left(\Pi(4m_{\chi}^2)\right)\left(2m_\chi^2+m_\sigma^2\right)|^2}{16\pi f^4(2m_{\chi}^2-m_{\sigma}^2)^2|4m_{\chi}^2-m_{\sigma}^2-i\text{Im}\left(\Pi(4m_{\chi}^2)\right)|^2}.\nonumber\\
\end{eqnarray}
Note that the second term in the parenthesis of Eq. (\ref{eq:wwann}), corresponding to the formation of longitudinal gauge boson modes, becomes proportional to $m_\chi^4/m_W^4$ in the limit $m_{\chi}\gg m_W$.  In this limit, the $m_W^4$ pre-factor is cancelled such that the overall cross section scales like $m_{\chi}^2/f^4$. 

In the expressions above $\Pi (p^2)$ is the 1PI self-energy insertion for the dilaton, which on-shell is related to the width via
  $m_{\sigma}\Gamma_{\sigma}=-\text{Im}\left(\Pi(m_{\sigma}^2)\right)$.  Note that we only include the imaginary part in our calculations. The real part (once properly renormalized) is expected to be a moderate correction to the existing real part of the propagator, which will result in small shifts to the precise shape of the contours presented below, but can not qualitatively change the results, as long as the NDA bound~(\ref{eq:NDA}) on the dilaton mass is obeyed. On the other hand properly incorporating the non-vanishing imaginary part can give significant shifts in the resulting cross sections especially close to the resonance. 
  
The total width of the dilaton is the sum of the partial widths to Higgs, quarks, massive gauge bosons and dark matter, which in the limit $m_{\sigma}\gg m_t$ is dominated by the decays to massive gauge bosons
\begin{eqnarray}
\Gamma_{\sigma}(\sigma \rightarrow WW) = \frac{m_W^4}{4\pi m_{\sigma}f^2}\sqrt{1-4\frac{m_W^2}{m_{\sigma}^2}}\left(2+\frac{(m_{\sigma}^2-2m_W^2)^2}{4m_W^4}\right).\label{eq:wwdec}
\end{eqnarray}
The processes $\chi\chi\rightarrow ZZ$ and $\sigma\to ZZ$ are obtained from Eqs.~(\ref{eq:wwann}-\ref{eq:wwdec}) by replacing $m_W$ by $m_Z$ and dividing by 2 to account for the phase space of identical final state particles. 
In Appendix \ref{app:B} we collect the contributions of the other channels to the dilaton decay width.  

\subsubsection*{Fermionic dark matter}

For fermionic dark matter, the annihilation channels have no s-wave contribution, thus the dominant contribution is a p-wave process which is suppressed by a factor of $v^2$.  We find 
\begin{eqnarray}
\sigma v(\chi\bar{\chi}\rightarrow WW) &\simeq& v^2 \frac{m_{\chi}m_W^4\sqrt{m_{\chi}^2-m_W^2}\left(2+\frac{(2m_{\chi}^2-m_W^2)^2}{m_W^4}\right)}{16\pi f^4|4m_{\chi}^2-m_{\sigma}^2-i\text{Im}\left(\Pi(4m_{\chi}^2)\right)|^2}
\end{eqnarray}
\begin{eqnarray}
\sigma v(\chi\bar{\chi}\rightarrow \sigma\sigma) &\simeq& v^2\Bigg[\frac{m_{\chi}^5\sqrt{m_{\chi}^2-m_{\sigma}^2}\left(9m_{\chi}^4-8m_{\sigma}^2m_{\chi}^2+2m_{\sigma}^4\right)}{24\pi f^4\left(16m_{\chi}^8-32m_{\chi}^6m_{\sigma}^2+24m_{\chi}^4m_{\sigma}^4-8m_{\sigma}^6m_{\chi}^2+m_{\sigma}^8\right)}\nonumber\\
&+& \frac{25m_{\chi}m_{\sigma}^4\sqrt{m_{\chi}^2-m_{\sigma}^2}}{128\pi f^4 |4m_{\chi}^2-m_{\sigma}^2-i\text{Im}\left(\Pi(4m_{\chi}^2)\right)|^2}\nonumber \\
&-&\frac{5m_{\chi}^3m_{\sigma}^2\sqrt{m_{\chi}^2-m_{\sigma}^2}(5m_{\chi}^2-2m_{\sigma}^2)}{48\pi f^4\left(4m_{\chi}^4-4m_{\sigma}^2m_{\chi}^2+m_{\sigma}^4\right)}{\rm Re}\left(\frac{1}{4m_{\chi}^2-m_{\sigma}^2-i\text{Im}\left(\Pi(4m_{\chi}^2)\right)}\right)  \Bigg].\nonumber\\
\end{eqnarray}

\subsubsection*{Vector dark matter}

For vector boson dark matter the annihilation is again dominated by s-wave processes: 
\begin{eqnarray}
\sigma v(\chi\chi\rightarrow WW) &\simeq& \frac{m_{\chi}m_W^4\sqrt{m_{\chi}^2-m_{W}^2}\left(2+\frac{(2m_{\chi}^2-m_W^2)^2}{m_W^4}\right)}{6\pi f^4|4m_{\chi}^2-m_{\sigma}^2-i\text{Im}\left(\Pi(4m_{\chi}^2)\right)|^2}\\
\sigma v(\chi\chi\rightarrow \sigma\sigma) &\simeq& \frac{m_{\chi}\sqrt{m_{\chi}^2-m_{\sigma}^2}}{144\pi f^4(2m_{\chi}^2-m_{\sigma}^2)^2|4m_{\chi}^2-m_{\sigma}^2-i\text{Im}\left(\Pi(4m_{\chi}^2)\right)|^2}\Big(708m_{\chi}^8\nonumber\\
&+&44m_{\sigma}^2m_{\chi}^2\text{Im}^2\left(\Pi(4m_{\chi}^2)\right)-28m_{\sigma}^4\text{Im}^2\left(\Pi(4m_{\chi}^2)\right)-1600m_{\sigma}^2m_{\chi}^6\nonumber\\
&+&1424m_{\chi}^4m_{\sigma}^4-576m_{\sigma}^6m_{\chi}^2+11m_{\sigma}^4\text{Im}^2\left(\Pi(4m_{\chi}^2)\right)+96m_{\sigma}^8\Big).
\end{eqnarray}

\subsection{Unitarity considerations}

We emphasize again that the $WW$ and $ZZ$ annihilation channels are important  because of the enhanced  contributions of the longitudinal modes.  Note that Ref.~\cite{Bai:2009ms} neglected these channels due to the suppression of the $W/Z$ couplings by $m_{W,Z}/f$. However as we have shown in the previous section, these factors are cancelled in the limit $m_{\chi}\gg m_Z$ due to the contributions of the longitudinal modes which grow with the CM energy/dark matter mass. 

For large DM mass, the gauge boson longitudinal modes might violate unitarity.  This is analogous to the unitarity violation in elastic WW scattering in the standard model without the Higgs.  However here the Higgs does not save unitarity.   Thus we will have a unitarity bound on the DM mass, related to the  built-in cutoff for the theory above which it is expected to be strongly coupled. One can estimate the unitarity bound on $m_\chi$  by considering the contribution of the longitudinal mode to the scattering amplitude in the large DM mass limit, given by  $\mathcal{M}\approx 2m_{\chi}^2/f^2$ for either scalar, fermion or vector DM.  The resulting s-wave partial wave amplitude $a_0\approx m_{\chi}^2/(16\pi f^2)$ satisfies  the unitarity bound $|\Re(a_0)|\leq 1/2$ if
\begin{eqnarray}
m_{\chi}\leq \sqrt{8\pi}f.\label{eq:unit}
\end{eqnarray}
This unitarity bound on $m_\chi$ is slightly more constraining than the NDA estimate for the cutoff $m_\chi\lesssim \Lambda_{NDA} = 4\pi f$.  A similar analysis for the annihilation to dilatons results in the same upper bound.

\subsection{The basic parameter space\label{sec:ParamSpace}}

\begin{figure}
\begin{center}
\includegraphics[width=.475\textwidth]{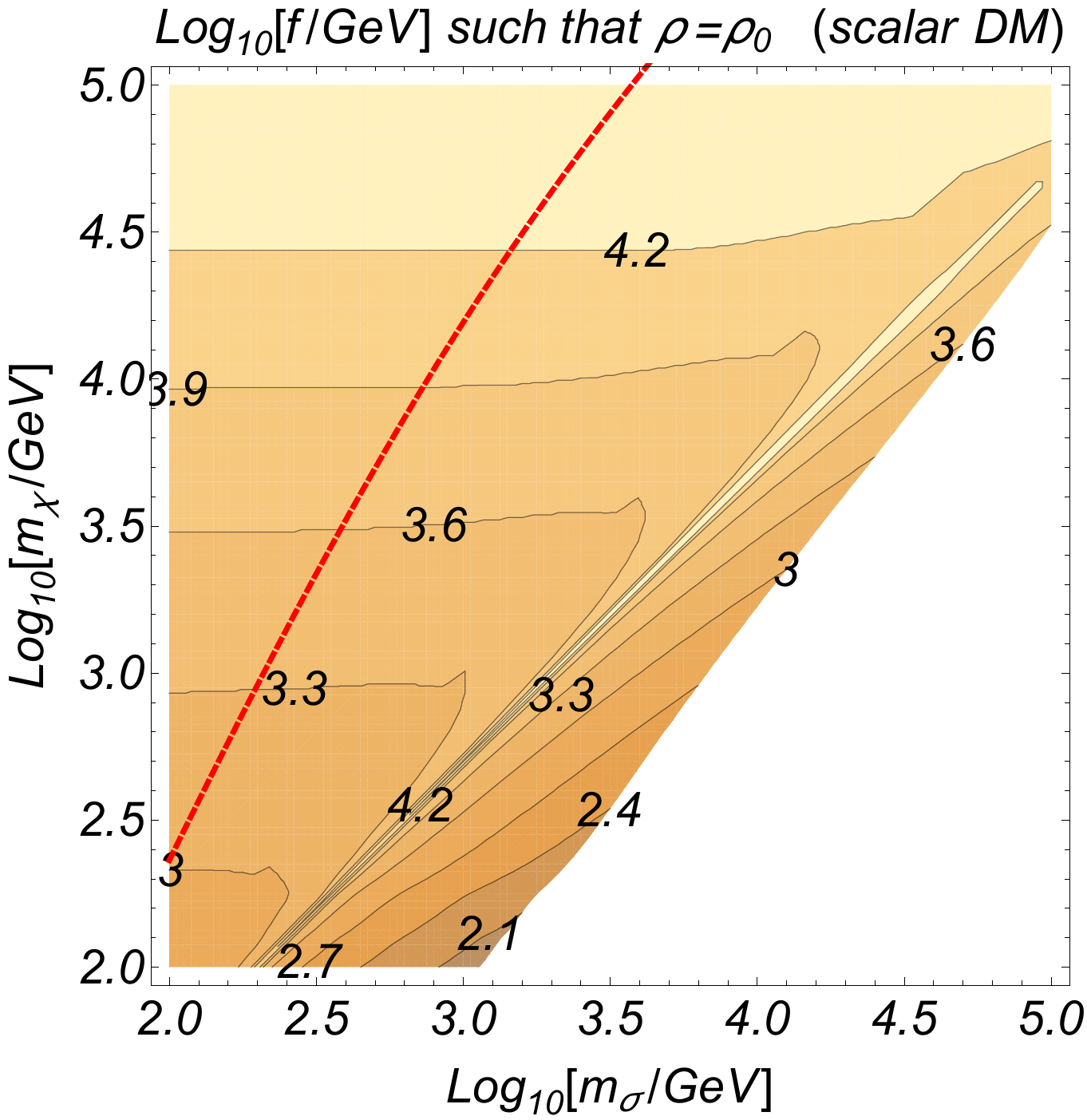}\;
\includegraphics[width=.475\textwidth]{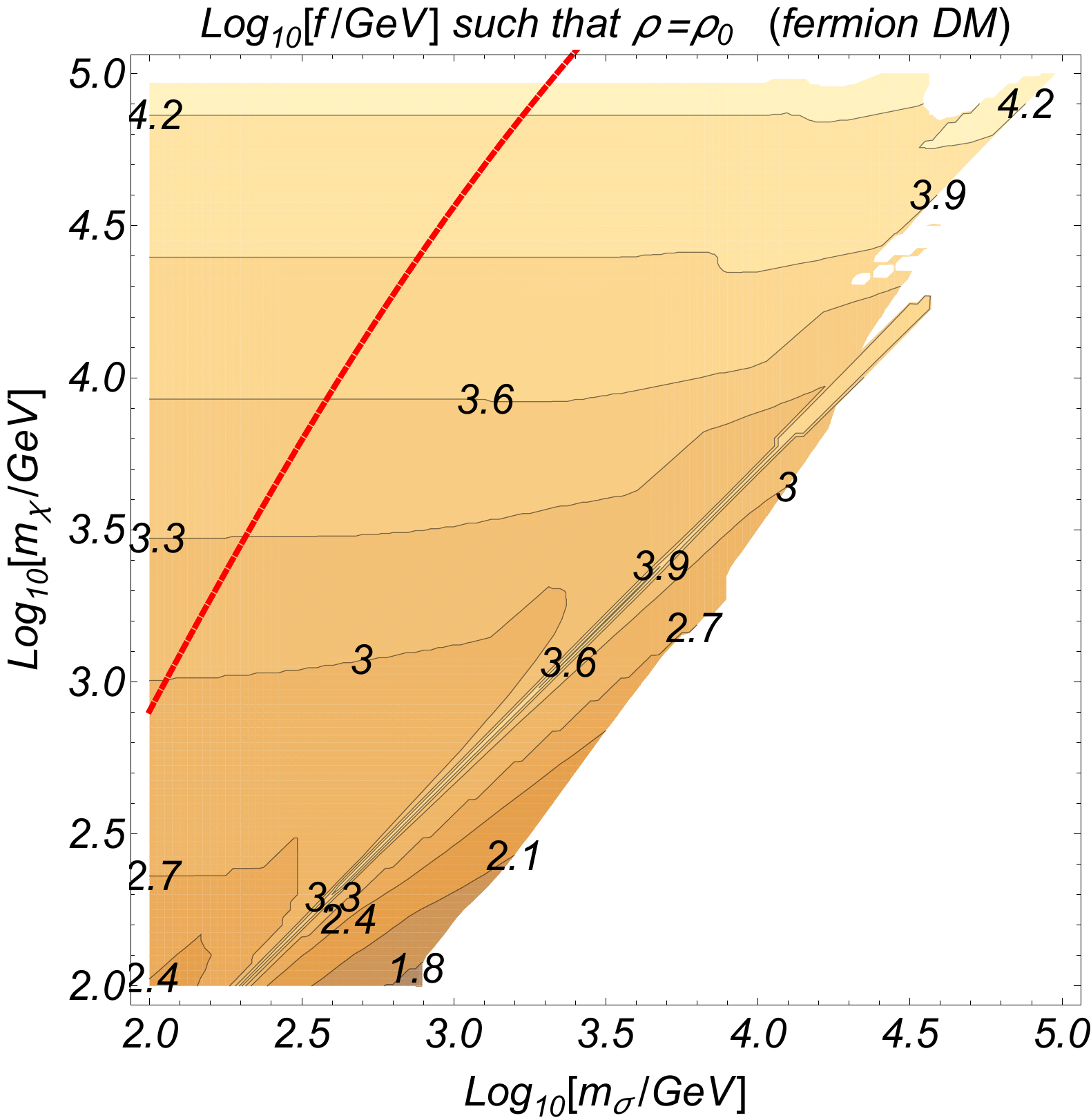}\;
\includegraphics[width=.475\textwidth]{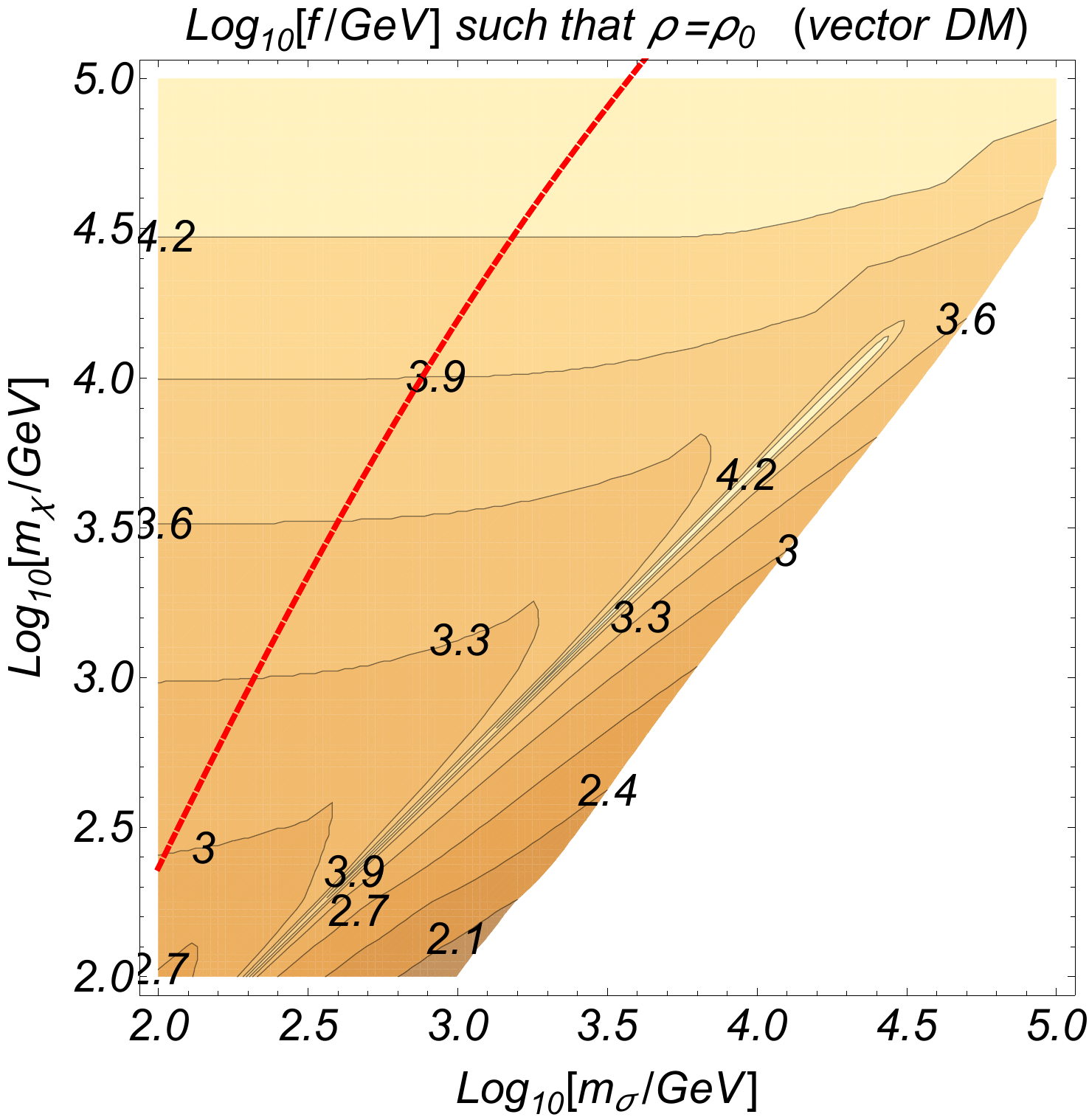}\;
\end{center}
\caption{Parameter space for scalar (top left), fermion (top right) and vector (bottom) dark matter with freeze-out mediated by dilaton exchange. The $x$ and $y$ axes correspond to the dilaton and dark matter mass, respectively. Contours show the value of the symmetry breaking scale $f$, that is required in order to obtain the observed dark matter relic density. In the blank region in the lower-right part of the plot, there is no real solution for $f$ that provides the correct relic density while satisfying Eq.~(\ref{eq:NDA}). Above the red dashed line $m_\sigma<f/10$, signaling some degree of fine-tuning. Note that the model-independent unitarity bound of Ref.~\cite{griest1990unitarity} implies $m_\chi\lesssim10^5$~GeV (see text).}
\label{fig:DMParamSpace}
\end{figure}

We now analyze the parameter space of the model that is compatible with the observed dark matter relic density. Fig.~\ref{fig:DMParamSpace} shows the available parameter space where the observed relic density can be reproduced by an appropriate choice of the symmetry breaking scale $f$. The top left, top right, and bottom panels show the results for scalar, fermion, and vector dark matter, respectively. The $x$- and $y$-axes correspond to the dilaton and dark matter mass, while the contours show the value of   $f$ that is required  to obtain the observed dark matter relic density. 
%%%%%%%%%%%%%%%%%%%%%%%%%%%%%%
%%%%%%%%%%%%%%%%%%%%%%%%%%%%%%

For concreteness, in the rest of this section we discuss the scalar dark matter case. We later summarize the results for fermion and vector dark matter. 
To understand the results shown in Fig.~\ref{fig:DMParamSpace} (top left, as we are focusing on the scalar example), we consider the different parametric regions in turn. Consider the case $m_\chi,m_\sigma\gg m_Z$, where annihilation to $WW,ZZ$ and, if kinematically allowed, $\sigma\sigma$ dominates. 
Assume first $m_\chi> m_\sigma$, corresponding to the upper-left region in Fig.~\ref{fig:DMParamSpace}. Here we have
\begin{eqnarray}
\langle\sigma v\rangle \approx \frac{m^2_{\chi}}{4\pi f^4}\approx3\times10^{-26}\left(\frac{f}{\rm 6~TeV}\right)^{-2}\left(\frac{m_\chi}{f}\right)^2\,{\rm cm^3/s}\;\;\;\;\;\;\;\;({\rm valid\;for\;m_\chi\gg m_\sigma}).\label{eq:svchiggsig}
\end{eqnarray}
Recall that relic abundance consistent with observations requires $\langle\sigma v\rangle \approx3\times10^{-26}$~cm$^3$/s, and that $\Omega_\chi h^2\propto\langle\sigma v\rangle^{-1}$, imposing the relation $m_\chi = f^2 /(6 {\rm TeV})$. Combining this with the unitarity bound  $m_\chi\sim\sqrt{8\pi}f$ obtained above, we find an upper bound $f< 30$~TeV. Violating this bound leads to DM annihilation cross section that is too small, and so DM relic density that is too high to match observations. A caveat in this derivation is that our dark matter particle may co-annihilate with extra particles in the dark sector. If this co-annihilation is efficient, due to some mass degeneracy in the dark sector and large cross sections, then it would relax the bound on $f$, allowing $f$ to be somewhat larger than $30$~TeV. Even taking this caveat into account, a rough bound $f\lesssim100$~TeV is still expected to hold. We note that this derivation of the bound on $f$ is compatible with the unitarity argument of~\cite{griest1990unitarity}, that showed that $m_\chi\lesssim100$~TeV is required in general from S-matrix unitarity (we update their early result here by using the currently measured DM relic density). Plugging the model-independent upper bound on $m_\chi$ from Ref.~\cite{griest1990unitarity} into Eq.~(\ref{eq:svchiggsig}), we obtain again $f\lesssim30$~TeV. The consistency between Eq.~(\ref{eq:unit}) and the unitarity bound of~\cite{griest1990unitarity} implies that Eq.~(\ref{eq:unit}) is satisfied throughout the parameter space shown in Fig.~\ref{fig:DMParamSpace}.

Next, consider the region with $m_\sigma\gg m_\chi$, so that the $\chi\chi\to\sigma\sigma$ channel is kinematically forbidden. This region corresponds to the lower-right part of Fig.~\ref{fig:DMParamSpace}. In this regime, and still assuming $m_\chi\gg m_Z$, one finds the following approximate form for the cross section:
\begin{equation}
\langle\sigma v\rangle \sim \frac{3 m_\chi^6}{\pi f^4 m_\sigma^4}
\approx2 \cdot 10^{-26}\,\left(\frac{m_\chi}{\rm 350 ~GeV}\right)^{6}\left(\frac{\rm TeV}{f}\right)^4 \left(\frac{\rm TeV}{m_\sigma}\right)^4\,{\rm cm^3/s}\;\;\;\;({m_\chi\gg m_W,\;m_\sigma\gg m_\chi}).\label{eq:svchilsig}\end{equation}
As one increases the dilaton mass $m_\sigma$ the symmetry breaking scale $f$ needs to decrease in order to keep the relic abundance fixed. However, one will very quickly need to lower $f$ below the value $m_\sigma/4\pi$, implying that we have left the regime of validity of our effective theory. Therefore most of the lower left region will be excluded based on this criterion. Of course the exact shape of the excluded region will be somewhat uncertain: it depends on the exact onset of strong coupling, and can also be slightly modified by strong co-annihilations in the dark sector. Nevertheless, even in this case we expect that the allowed region would shift only slightly. 

The resonance at $m_\sigma=2m_\chi$ is clearly visible in Fig.~\ref{fig:DMParamSpace}. 
The approximate expression of the cross section close to the resonance region is 
\begin{eqnarray}
\langle \sigma v \rangle \sim \frac{3 m_\chi^6}{\pi \left[ (\Delta m)^4 f^4 +\frac{9 m_\chi^8}{4\pi^2} \right]}\ ,
\end{eqnarray}
where $\Delta m^2 = 4 m_\chi^2 - m_\sigma^2$, measuring the deviation from the exact location of the resonance.  In this region (but above the blank region corresponding to Eq.~(\ref{eq:svchilsig})), a large value of $f$ is required to reduce the otherwise too high annihilation cross section. Note, that once $m_\chi \sim 40$ TeV the cross section falls below the observed value even without a contribution from the resonance. Above those masses one does not expect any more resonant behavior, which is indeed what is reflected in Fig.~\ref{fig:DMParamSpace}. We note that numerical resolution affects the size of $f$ that is displayed in Fig.~\ref{fig:DMParamSpace} exactly on the resonance line, as $f\to\infty$ for $\Delta m^2\to 0$. Of course, living exactly on resonance corresponds to an extremely fine-tuned parametric set-up. Note that beyond the mere parametric fine-tuning, another issue here is that $f\gg m_\sigma$ would imply dynamical fine-tuning as well.

We conclude the discussion of the scalar DM case by considering the scenario proposed in Ref.~\cite{Bellazzini:2012vz}, that entertained the possibility of having the newly discovered Higgs-like particle itself be the dilaton. For the dilaton to mimic the Higgs, we must have $m_\sigma\approx m_h=126$~GeV and $f\approx v=246.2$~GeV. For these values of $m_\sigma$ and $f$, we find that the dark matter mass that is needed for correct relic abundance is $m_\chi\approx52$~GeV if the dark matter is a scalar. The leading annihilation channels at this value of $m_\chi$ are to bottom and charm quarks and tau leptons. Larger values of $m_\chi$ result in relic abundance that is too low, while lower values of $m_\chi$ give a too-high relic abundance. This means that $m_\chi\approx52$~GeV is an upper bound for scalar dark matter mass in our framework in the Higgs-like dilaton scenario. As we show in Sec.~\ref{sec:dd}, such a low scalar dark matter mass is excluded by direct detection limits. Similar results are obtained for the case of fermion and vector DM, as presented in the second and third plots in Fig.~\ref{fig:DMParamSpace}. The higgs-like dilaton scenario would require fermion dark matter of 61 GeV, or vector DM of 56 GeV. As we will see both of these cases are excluded by the direct detection bounds. 

Finally, note that in part of the parameter space depicted in Fig.~\ref{fig:DMParamSpace} the DM annihilation cross section receives large non-perturbative corrections at low center of mass velocities (Sommerfeld enhancement). In our model, at large DM mass when the effective coupling $m_\chi/f$ is not far from the perturbativity limit, the effect induces a sizable correction to the relic abundance calculation.
We compute the Sommerfeld enhancement in Sec.~\ref{sec:id} and include it in a simplified form in the calculation of Fig.~\ref{fig:DMParamSpace}, by rescaling the tree-level annihilation cross section by the Sommerfeld enhancement factor at relative DM velocity $v=0.3$, corresponding roughly to the thermal freeze-out kinematics. In most of the parameter space, corresponding to perturbative coupling $(m_\chi/f)^2/4\pi\ll1$, the correction to the derived value of $f(m_\chi,m_\sigma)$ fixed by the  relic abundance requirement is insignificant~\footnote{In fine-tuned regions of the parameter space, where the Sommerfeld effect hits a resonance, DM annihilation re-coupling can significantly affect the relic abundance calculation~\cite{Feng:2010zp}. We ignore this effect here and comment about it in Sec.~\ref{sec:id}.}.

%%%%%%%%%%%%%%%%%%%%%%%%%%%%%%%
\section{Direct Detection} \label{sec:dd}

Having defined the parameter space of the theory that reproduces the correct relic abundance, we now study direct detection constraints. For direct detection we need to consider the elastic cross section of a dark matter particle that scatters off a nucleon.  The dilaton interacts with quarks $q$ and the gluons $G^{a\mu\nu}$ inside a nucleon~\cite{Fox:2011pm,Junnarkar:2013ac}. Thus the relevant part of the dilaton effective Lagrangian is 
\begin{equation}
{\cal L} \supset  -\sum_q\frac{\sigma}{f} (1+\gamma_q) m_q q \bar{q} +\frac{\alpha_s}{8\pi f} c_G G^2\,.
\end{equation}
To estimate the anomalous dimension for quarks, one can consider the corresponding warped extra dimensional models where the anomalous dimension is determined~\cite{CHL} by $1+\gamma = c_L-c_R$, where $c_{L,R}$ are the bulk fermion mass parameters. For typical warped fermion scenarios we find for example $\gamma_s \sim 0.16$, which we  neglect in the bounds below.

Taking the matrix element between nucleon states yields the effective nucleon-dilaton Lagrangian
\begin{eqnarray}
\mathcal{L}_{\sigma nn} &=& y_{n} \sigma n\bar{n}
\end{eqnarray}
where the coefficient $y_n$ is determined by the $f^n_q,R^n$ hadronic matrix elements:
\begin{equation}
y_{n} \equiv  - \sum_q f_q^n \frac{m_n}{f} + R^n\frac{ c_G}{8\pi f}\ .
\end{equation}
For these matrix elements we use the values from~\cite{Junnarkar:2013ac,Bai:2010hh,Ellis:2008hf}: 
\begin{eqnarray}
f_{q}^n&=&\langle n|\bar{q}q|n\rangle \frac{m_q}{m_n}\nonumber\\
f_{u}^n &\simeq& f_{d}^n \simeq 0.022 \nonumber\\
f_{s}^n &\simeq& 0.043\nonumber\\
f_{c}^n &\simeq& 0.0814\nonumber\\
f_{b}^n &\simeq& 0.0785\nonumber \\
f_{t}^n &\simeq& 0.0820\nonumber \\
R^n&=&\alpha_s \langle n| G^a_{\mu\nu}G^{a\mu\nu}|n\rangle \simeq -2.4\text{GeV}
\end{eqnarray}
With this effective interaction the scattering cross section between dark matter and nucleons is given by
\begin{eqnarray}
\sigma_{\chi,n} &\approx&\frac{ y_{n}^2 }{\pi} \left(\frac{m_{\chi}}{f}\right)^2 \frac{m_n^2}{m_{\sigma}^4}
\end{eqnarray} 
for either scalar, fermionic or vector dark matter. 

\begin{figure}
\hspace*{-0.5in}
\begin{subfigure}[b]{0.6\textwidth}
			\includegraphics[width=\textwidth]{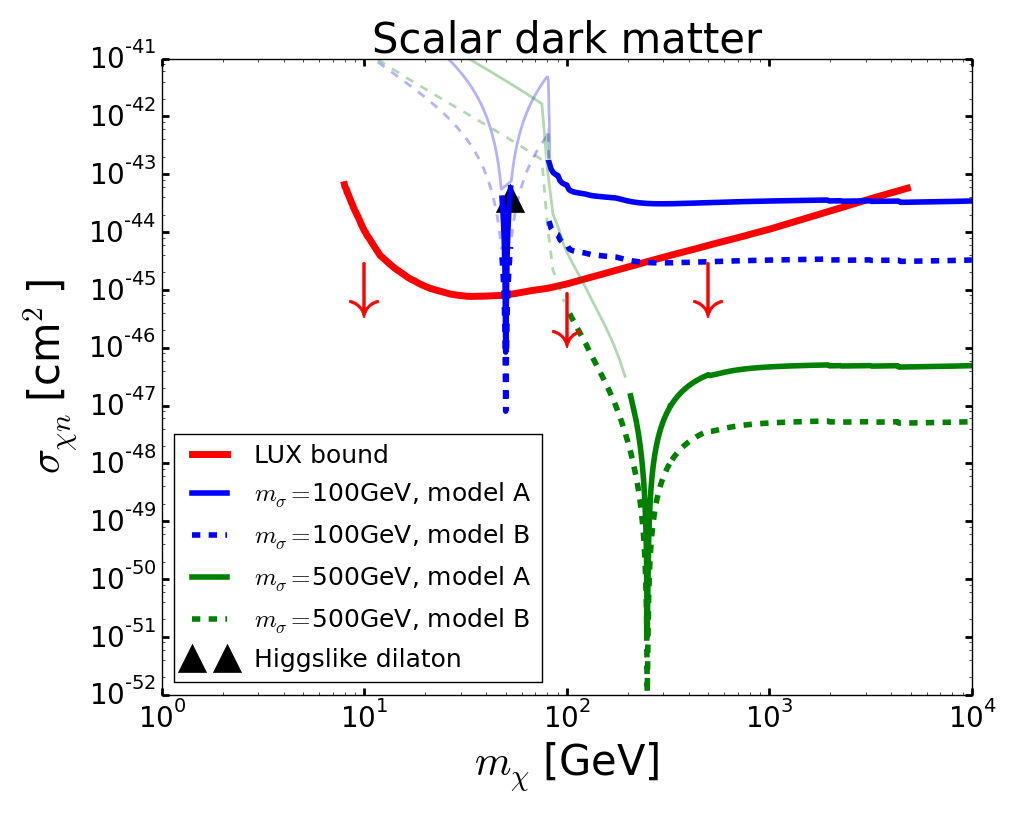}
		\end{subfigure}
		\begin{subfigure}[b]{0.6\textwidth}
			\includegraphics[width=\textwidth]{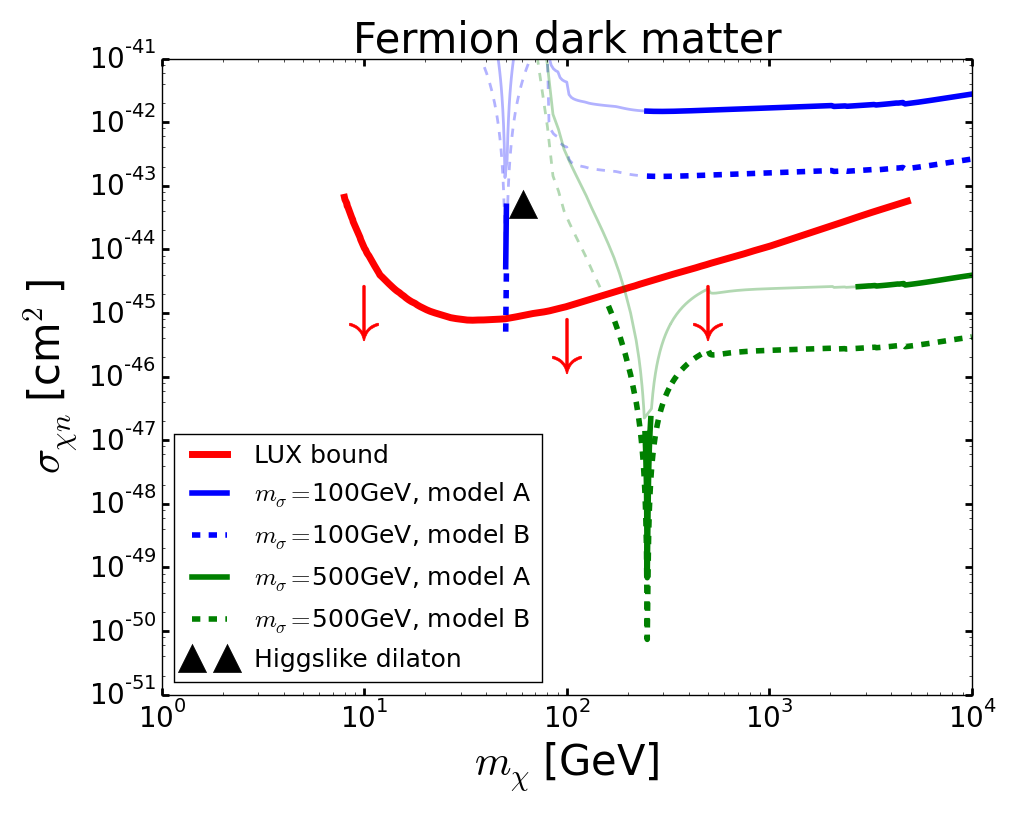}
		\end{subfigure}
\begin{center}
\includegraphics[width=.6\textwidth]{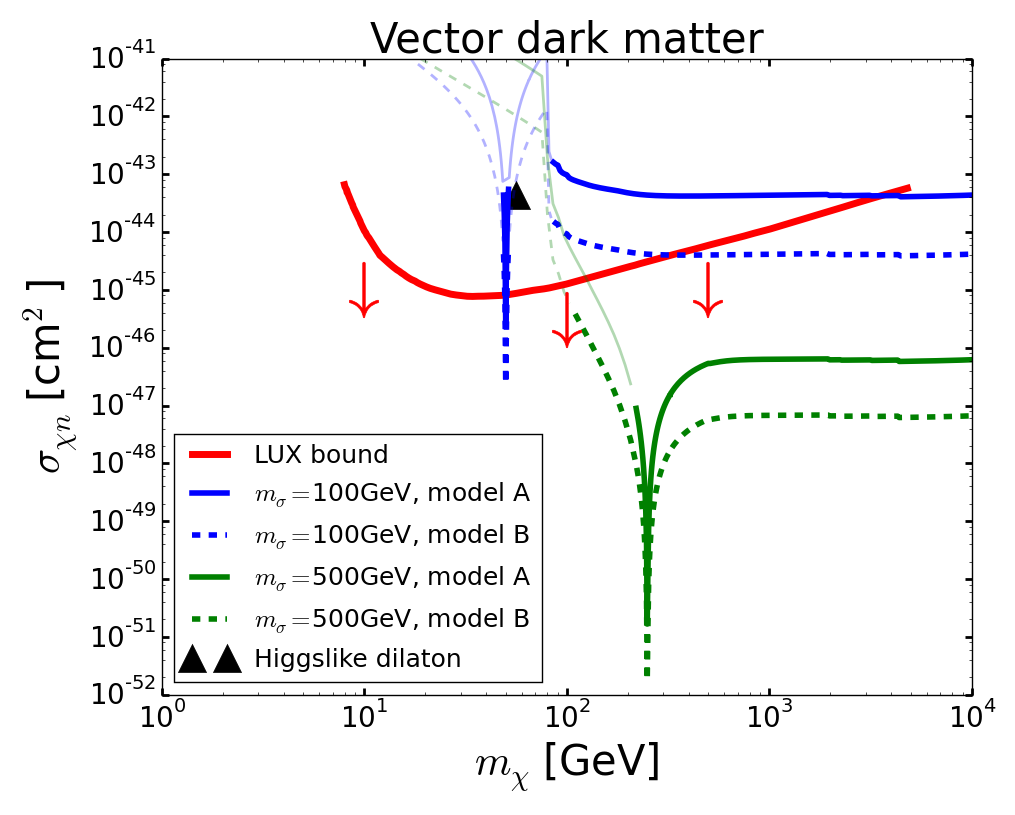}\;
\end{center}
\caption{Nucleon-dark matter elastic cross section as a function of dark matter mass. The red arrows point towards the non-excluded region.  The lighter portion of the curves are already excluded by bounds from collider experiments searching for a dilaton.}
\label{fig:dd}
\end{figure}

Fixing the scale $f$ for given $m_\sigma$ and $m_\chi$ to match the relic abundance, we plot the DM-nucleon elastic scattering cross section as a function of the dark matter mass for a few dilaton mass values.  The results are illustrated on Fig.~\ref{fig:dd} along with the recent direct detection constraints from the LUX experiment \cite{Akerib:2013tjd}. We have also  included the effects of the collider bounds on the dilaton from the LHC and other machines (see Appendix \ref{app:C}).  These plots show that most of the parameter space is currently allowed both by the dark matter direct detection experiments and also by the collider constraints, as long as $m_{\sigma}\gtrsim 200$ GeV.  

As discussed in Sec.~\ref{sec:ParamSpace}, for $m_{\chi}\gg m_t$ and away from the resonance the annihilation cross section is proportional to $m_{\chi}^2/f^4$.  Moreover, since $y_n\propto 1/f$, we can see that the elastic scattering cross section is proportional to the same combination $m_{\chi}^2/f^4$. Thus in the appropriate regime  the elastic cross section will be  independent of the dark matter mass, as can be seen in Fig.~\ref{fig:dd}.  

%%%%%%%%%%%%%%%%%%%%%%%%%%%%%%%
\section{Sommerfeld Enhancement and Indirect Detection} \label{sec:id}

We now consider the prospects for indirect detection of dark matter annihilation via gamma ray and cosmic ray antiproton flux measurements~\footnote{Additional constraints can be derived from neutrino experiments. These constraints are typically weaker than those arising from gamma ray and antiproton data (see e.g.~\cite{Agashe:2009ja} for discussion of the $\nu$ flux in the context of a related model) and we do not consider them in this work. Under specific cosmic ray propagation model assumptions, constraints can also be derived from the high energy positron flux. In comparison to the $\bar p$ calculation, however, the theoretical uncertainties for $e^+$ are larger as the results depend crucially on the cosmic ray propagation time in the Galaxy that dictates the amount of $e^+$ radiative energy loss~\cite{Blum:2010nx}, and so we do not consider $e^+$ constraints in this work.}. 
We limit the discussion to the case in which the DM $\chi$ is a real scalar field. We expect similar results for the vector DM case; the fermion DM case will not have significant cosmic ray signatures as its annihilation is p-wave suppressed in the small virial velocity of the Milky Way and its dwarf satellite galaxies. 

\begin{figure}
\begin{center}
\includegraphics[width=.675\textwidth]{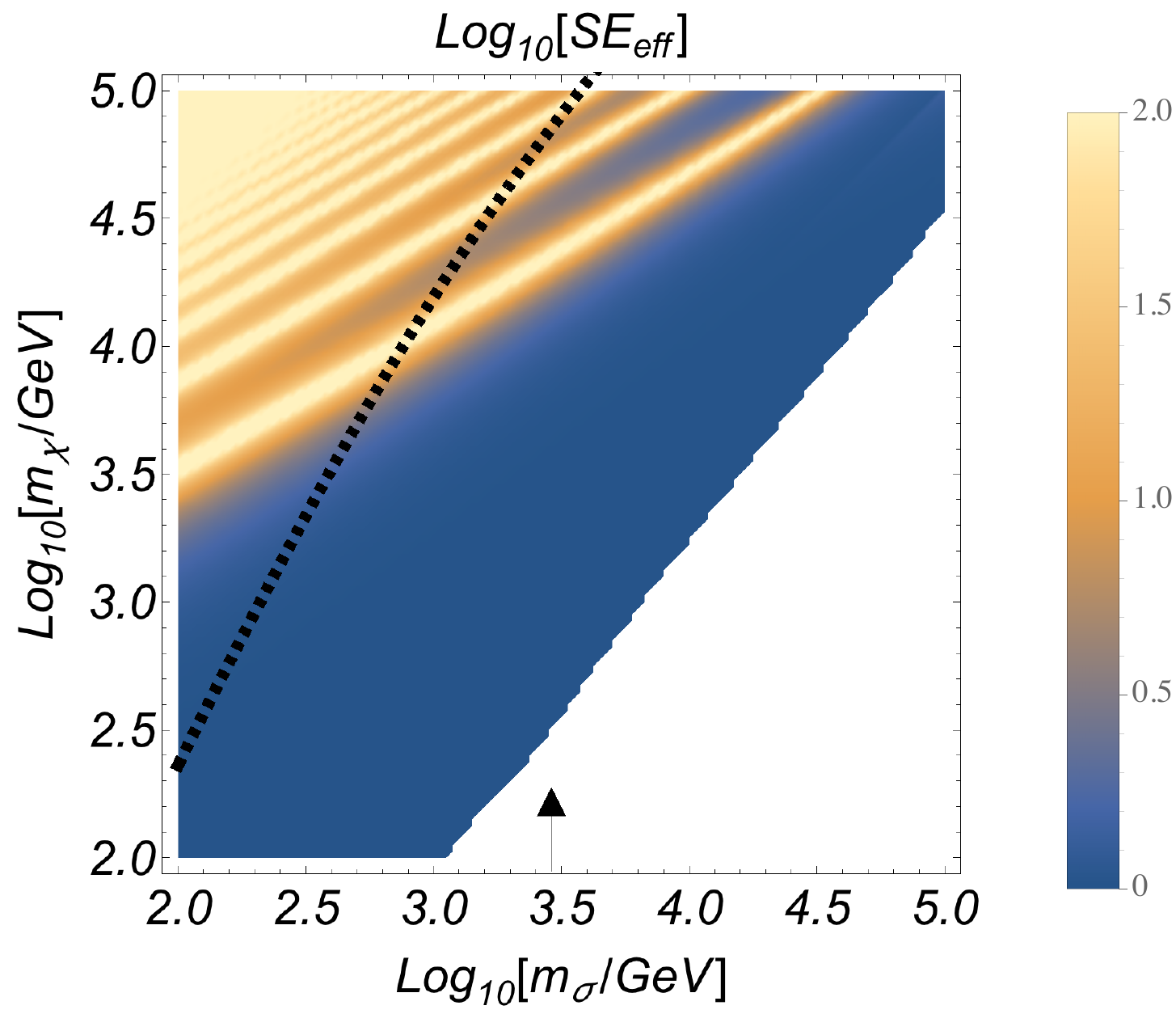}\;
\includegraphics[width=.65\textwidth]{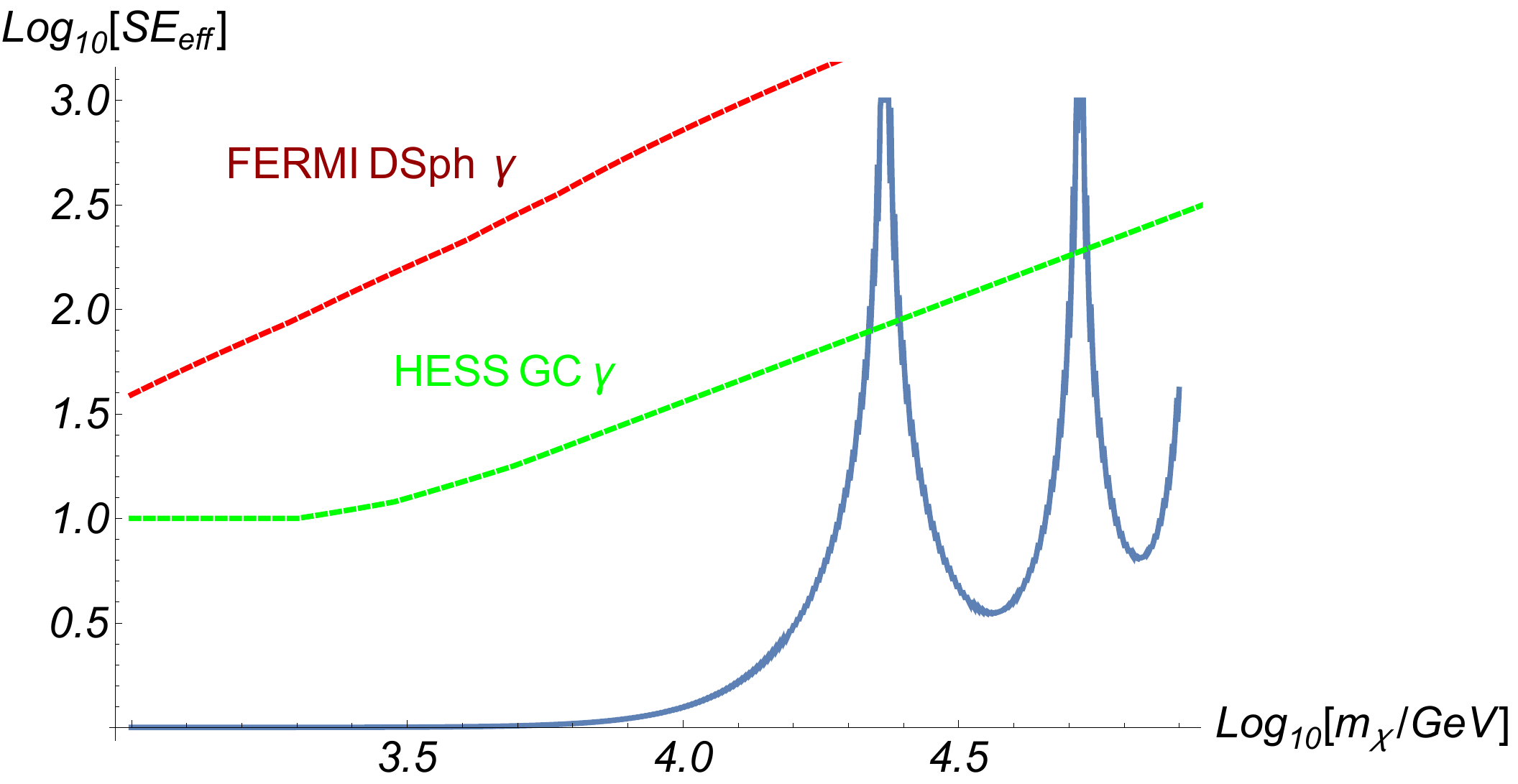}
\end{center}
\caption{Top panel: Sommerfeld enhancement factor (SE) in the $\{m_\sigma,m_\chi\}$ plane. Above the dashed line $m_\sigma<f/10$, indicating fine-tuning. Bottom panel: SE vs. dark matter mass, fixing the dilaton mass to $m_\sigma=3$~TeV  (marked on top panel with an arrow). The region above the red and green dashed lines is excluded by FERMI and HESS gamma ray observations (the latter depend strongly on assumptions regarding the DM distribution in the Galaxy; see Sec.~\ref{ssec:gr}). The dark matter particles' relative velocity today is set to $v=10^{-3}$.}
\label{fig:SEplot}
\end{figure}

The parameter space of interest for the model includes the regime where $m_\chi> m_\sigma$. In this regime, dilaton exchange produces an attractive Yukawa potential $-\frac{\alpha}{r}e^{-m_\sigma r}$, with $\alpha=\frac{m_\chi^2}{4\pi f^2}$, that affects the dark matter annihilation process giving rise to Sommerfeld enhancement (SE; see e.g.~\cite{Hisano:2003ec, Hisano:2004ds}) that needs to be taken into account in the indirect detection estimates. 
In the top panel of Fig.~\ref{fig:SEplot} we plot the effective SE factor (denoted $SE_{eff}$) in the $\{m_\sigma,m_\chi\}$ plane, fixing the value of the scale $f$ at each point to match the observed dark matter relic abundance. We define $SE_{eff}$ as the value of the SE today in the Galactic halo, normalized to its value during DM freeze-out when $v\sim0.3$. 
In our calculation we use an approximate formula for the SE factor~\cite{Cassel:2009wt,Feng:2010zp,Slatyer:2009vg},
\begin{eqnarray}
SE\approx\frac{\pi}{\epsilon_v}\frac{\sinh\left(\frac{12\epsilon_v}{\pi\epsilon_\phi}\right)}{\cosh\left(\frac{12\epsilon_v}{\pi\epsilon_\phi}\right)-\cos\left[2\pi\sqrt{\frac{6}{\pi^2\epsilon_\phi}-\left(\frac{12\epsilon_v}{\pi\epsilon_\phi}\right)^2}\right]},\label{eq:SEan}
\end{eqnarray}
where $\epsilon_v\equiv\frac{v}{2\alpha}=\frac{2\pi vf^2}{m_\chi^2}$ and $\epsilon_\phi\equiv\frac{m_{\sigma}}{\alpha m_{\chi}}=\frac{4\pi m_\sigma f^2}{m_\chi^3}$. We set the value of the dark matter particles' relative velocity to $v=10^{-3}$, appropriate for annihilation in the Galactic halo.
We have verified that the approximation above reproduces the full Sommerfeld calculation to a good accuracy. 
%%%%%%%%%%%%%%%%%%%%%%%%%%%%%%
%%%%%%%%%%%%%%%%%%%%%%%%%%%%%%

The top panel of Fig.~\ref{fig:SEplot} shows that for DM mass above a few TeV, large values of the SE factor are possible with $SE_{eff}>10^2$ in resonance regions. As we show below, this result may have interesting implications -- striking indirect detection signatures are possible if the model happens to live at an SE resonance. However, resonant SE is limited to fine-tuned regions in the parameter space. To illustrate this point, in the bottom panel we plot the value of SE vs. the DM mass fixing $m_\sigma=3$~TeV (corresponding to a vertical slice through the center of the top panel, marked by an arrow). For generic parameter configuration the effective SE factor is modest, and only grows above $10^2$ near resonances and for extremely heavy DM mass, close to the unitarity limit where our calculation breaks down. Note that we truncate the value of $SE_{eff}$ at $10^3$ in resonance peaks. As the resonance regions are fine-tuned, this has limited impact on our analysis. According to the analysis of~\cite{Feng:2010zp}, the relic abundance is depleted at the tip of these SE resonances due to chemical re-coupling of DM at low redshifts, an effect that we do not include here and that would reduce the value of $SE_{eff}$. In addition, the low velocity divergence of the SE at the resonance tip should be regulated by bound-state decay that would also suppress the peak SE.  

In Secs.~\ref{ssec:pb} and~\ref{ssec:gr} below we calculate antiproton and gamma ray constraints on the model. For antiprotons we adopt a conservative model-independent approach to the problem of cosmic ray propagation, and provide some extra details to explain our method. The summary of our results is that the bulk of the parameter space of Fig.~\ref{fig:SEplot} (or equivalently Fig.~\ref{fig:DMParamSpace}) is allowed by current constraints. This is not a surprise: much of the parameter space consistent with the DM relic density corresponds to rather heavy $m_\chi$ at the several TeV, where current indirect searches do not yet constrain the thermal relic cross section. Indirect detection constraints do exclude, or make promising predictions for, the near-resonant SE regions seen in Fig.~\ref{fig:SEplot}. If one accepts the assumption of a cusp DM density profile in the Milky Way Galactic Center, for example, then HESS gamma ray data already excludes much of the parameter region in the upper-left corner of the top panel of Fig.~\ref{fig:SEplot}.

\subsection{Antiprotons}\label{ssec:pb}
The PAMELA satellite experiment reported a measurement of the high energy antiproton flux in interstellar space, extending up to 350~GeV~\cite{Mayorov:2013coa}. The PAMELA measurement is consistent with model-independent calculations of the antiproton flux expected due to fragmentation of high energy primary cosmic ray nuclei on ambient interstellar gas in the Galaxy~\cite{Katz:2009yd}. 

Following Ref.~\cite{Agashe:2009ja}, we derive a bound on the antiproton production in dark matter annihilation by imposing that the dark matter annihilation source of antiprotons in the local Galactic gas disc does not exceed the source due to the astrophysical production, in the energy range covered by the current measurements. The bound derived in this manner is independent of modeling assumptions regarding the propagation of charged cosmic rays in the Galaxy. The bound is conservative because it does not include the possible additional contribution of DM annihilation in the cosmic ray halo that may extend well above and below the gas disc.  

\begin{figure}
\begin{center}
\includegraphics[width=.475\textwidth]{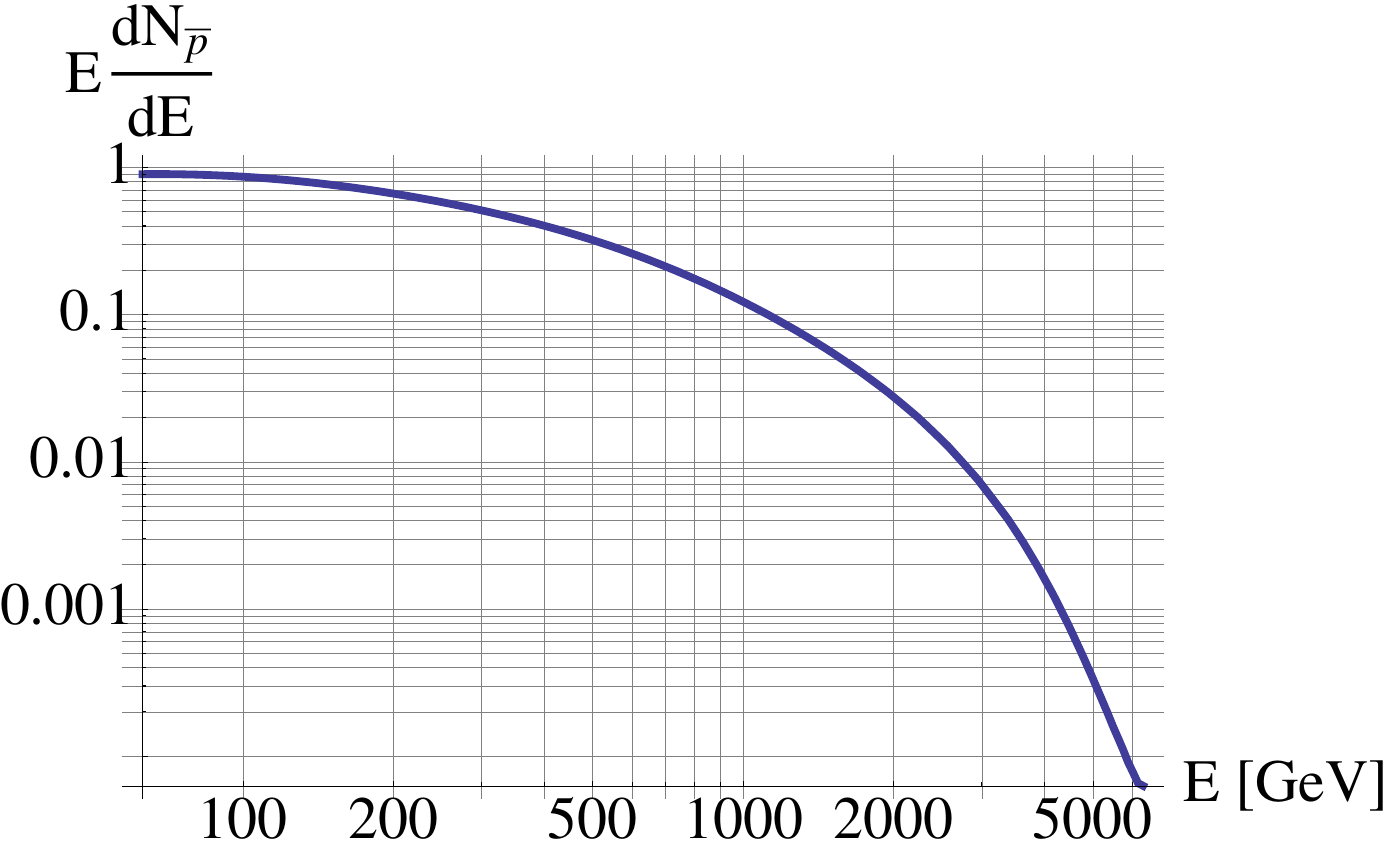}\;
\includegraphics[width=.475\textwidth]{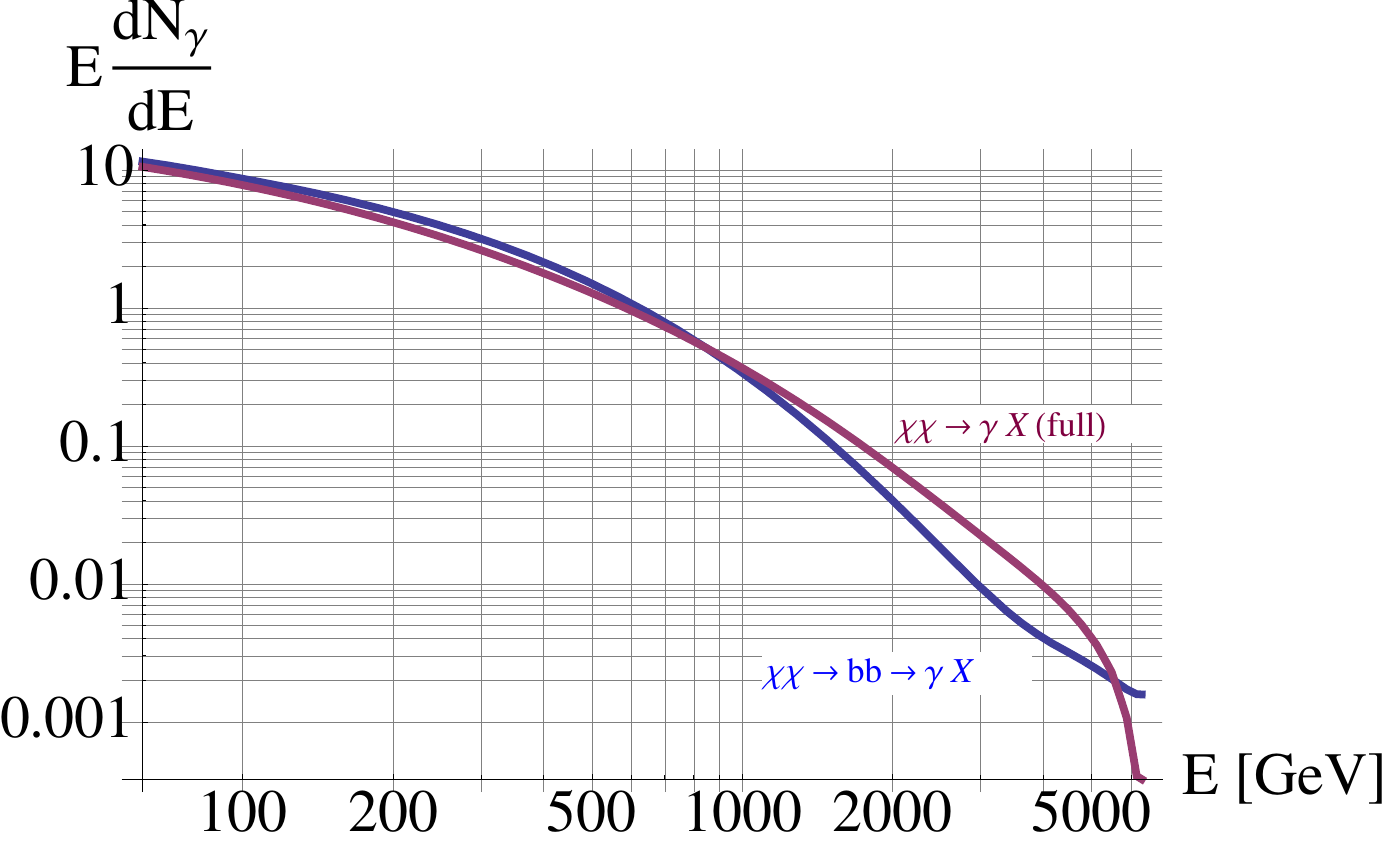}
\end{center}
\caption{Left: differential $\bar p$ spectrum per DM annihilation, computed for $m_\chi=6.3$~TeV, $m_\sigma=427$~GeV, and $f=6.2$~TeV. Right: same for the gamma ray spectrum; the purple line shows the spectrum from the full annihilation process including all dominant partial channels ($\chi\chi\to WW,ZZ,tt,\sigma\sigma,...$), while the blue line shows the spectrum due to $\chi\chi\to bb$ alone.}
\label{fig:dNdE}
\end{figure}

The injection rate density of antiprotons due to DM annihilation is given by
\begin{eqnarray}  &Q_{\bar p,DM}(E)&=\frac{1}{2}n_\chi^2\langle\sigma v\rangle\,\frac{dN_{\bar p}}{dE} \approx 5\times10^{-36}{\rm cm^{-3}s^{-1}GeV^{-1}} \times \nonumber \\
&& \left(\frac{\rho_\chi}{0.4~{\rm GeVcm^{-3}}}\right)^2\left(\frac{\langle\sigma v\rangle}{3\times10^{-26}~{\rm cm^3s^{-1}}}\right)\left(\frac{m_\chi}{1~\rm TeV}\right)^{-3}\left(m_\chi\frac{dN_{\bar p}}{dE}\right).
\label{eq:QpbDM}
\end{eqnarray}
Here, $\rho_\chi = m_\chi n_\chi\approx0.4$~GeV$\,$cm$^{-3}$ is the DM mass density in the local halo and $\frac{dN_{\bar p}}{dE}$ is the differential antiproton spectrum per annihilation event. To compute $\frac{dN_{\bar p}}{dE}$ we use the code provided in Ref.~\cite{Cirelli:2010xx}, that directly produces the differential $\bar p$ spectrum for the channels $\chi\chi\to WW,ZZ,hh,t\bar t$ accounting for the decay and hadronization of the intermediate unstable states. To include the contribution of $\chi\chi\to\sigma\sigma$ we proceed in two steps. First we use Ref.~\cite{Cirelli:2010xx} to calculate the $\bar p$ spectrum arising in the dilaton rest frame due to dilaton decay; define this spectrum by $\left[\frac{dN_{\bar p}}{dE}(E)\right]_{\sigma\to\bar pX}$. We then convolve the dilaton decay $\bar p$ spectrum with the isotropic boost factor of the $\sigma$ in the DM annihilation center of mass frame, obtaining 
\be
\left[\frac{dN_{\bar p}}{dE}(E)\right]_{\chi\chi\to\sigma\sigma}=\frac{1}{\gamma_\sigma\beta_\sigma}\int_{\beta_\sigma^{-1}-1}^{\beta_\sigma^{-1}+1}\frac{dx}{x}\left[\frac{dN_{\bar p}}{dE}\left(\frac{E}{x\gamma_\sigma\beta_\sigma}\right)\right]_{\sigma\to\bar pX}\ee
where $\gamma_\sigma=m_\chi/m_\sigma$ and $\beta_\sigma=\sqrt{1-\gamma_\sigma^{-2}}$.
We neglect DM annihilation into gluons, since the branching fraction of annihilation to this state is small compared to that of annihilation to quarks and massive gauge bosons. In the left panel of Fig.~\ref{fig:dNdE} we plot the differential flux of $\bar p$ from DM annihilation with $m_\chi=6.3$~TeV, $m_\sigma=427$~GeV, and $f=6.2$~TeV reproducing the observed DM relic abundance.
%%%%%%%%%%%%%%%%%%%%%%%%%%%%%%
%%%%%%%%%%%%%%%%%%%%%%%%%%%%%%

The injection rate density due to primary cosmic rays colliding with interstellar gas in the disc is~\cite{Katz:2009yd}
\begin{eqnarray} Q_{\bar p,CR}(E)&\approx&
8.4\times10^{-33}{\rm cm^{-3}s^{-1}GeV^{-1}} \times \nonumber \\
&& \left(\frac{E}{100~{\rm GeV}}\right)^{-2.8}\left[1-0.22\log_{10}^2\left(\frac{E}{500~{\rm GeV}}\right)\right]\,\frac{J_p(1~{\rm TeV})}{J_{p,0}(1~{\rm TeV})},
\label{eq:QpbCR}
\end{eqnarray}
where $J_p(1~{\rm TeV})$ is the local proton flux sampled at $E=1~{\rm TeV}$ and scaled to the measured value $J_{p,0}(1~{\rm TeV})\approx8\times10^{-9}~{\rm GeV^{-1}cm^{-2}s^{-1}sr^{-1}}$. The uncertainties in the derivation of Eq.~(\ref{eq:QpbCR}) are at the $\sim$50\% level. 
Our conservative bound on the DM annihilation rate amounts to imposing that the ratio $Q_{\bar p,CR}(E)/Q_{\bar p,DM}(E)$ is larger than unity for $E$ in the range 10-300~GeV.

The basic result we find is that the model survives our antiproton constraint by a large margin, unless it lives right on top of an SE resonance. If the model is near an SE resonance, then a detectable rise in the antiproton flux at high energy is predicted. For DM mass below about $\sim10$~TeV, the rise would be in tension with currently available $\bar p$ data and the model is observationally disfavored (again, only the region near an SE resonance, as seen in Fig.~\ref{fig:SEplot}). For $m_\chi\gtrsim10$~TeV, though, the rise in the $\bar p$ flux sets in at high energy with only a moderate effect in the energy range where current data exists. In this case, improved high energy cosmic ray measurements expected in the near future~\cite{Kounine:2012ega} may detect the model in the $\bar p$ flux. 

\begin{figure}
\begin{center}
\includegraphics[width=0.65\textwidth]{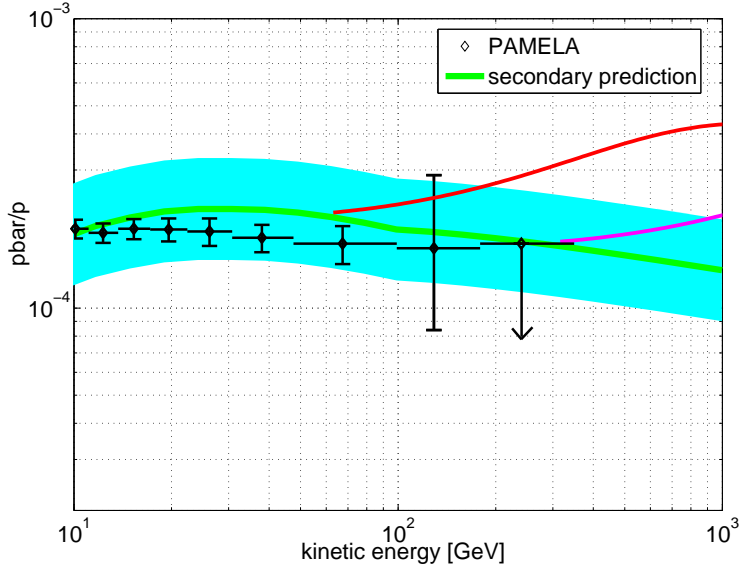}
\end{center}
\caption{Antiproton flux with DM annihilation at a Sommerfeld factor resonance. Data points and green curve denote PAMELA data and secondary astrophysics prediction, respectively. Red and magenta curves give a lower estimate of the $\bar p$ flux with DM annihilation for the model parameter point with $\{m_\chi=6.3~{\rm TeV}, m_\sigma=300~{\rm GeV}\}$ and $\{m_\chi=31~{\rm TeV}, m_\sigma=4.7~{\rm TeV}\}$, respectively, where the SE factor is $SE_{eff}\approx10^3$.}
\label{fig:pbar}
\end{figure}

We illustrate these findings in Fig.~\ref{fig:pbar} where we plot the expected antiproton flux in our model near an SE resonance for two chosen points. The data points (last one being an upper bound) and the green curve denote the PAMELA data and the secondary astrophysics prediction, respectively. The red and magenta curves give an estimate of the antiproton flux that would occur for the parameter points $\{m_\chi=6.3~{\rm TeV}, m_\sigma=300~{\rm GeV}\}$ and $\{m_\chi=31~{\rm TeV}, m_\sigma=4.7~{\rm TeV}\}$, respectively, where the effective SE factor is $SE_{eff}\approx10^3$ (fixing $f$ to obtain the observed DM relic abundance).  

Above we chose tuned points with large $SE_{eff}$ to illustrate the possible $\bar p$ signal; as mentioned earlier, this large $SE$ near the resonance peak can be damped somewhat by a more careful treatment of the relic abundance. However, we stress that the DM-induced signal depends on unknown cosmic ray propagation features. The red and magenta curves in Fig.~\ref{fig:pbar} should be considered as a robust {\it lower bound} on the DM-induced flux. Considering disc+halo diffusion models~\cite{Ginzburg:1990sk}, for example, the actual flux could be as high as a factor of $\sim100$ above the result we show here\footnote{See App.~B of Ref.~\cite{Agashe:2009ja} for a detailed discussion.}. A future detection of the model through cosmic ray $\bar p$ is therefore conceivable also away from SE resonance peaks. 
%%%%%%%%%%%%%%%%%%%%%%%%%%%%%%
%%%%%%%%%%%%%%%%%%%%%%%%%%%%%%

\subsection{Gamma Rays}\label{ssec:gr}
The FERMI gamma ray telescope reported limits on DM annihilation based on a stacking analysis of dwarf spheroidal galaxies~\cite{Ackermann:2013yva}. The analysis is relatively insensitive to the assumed DM mass distribution in the target galaxies. Ref.~\cite{Ackermann:2013yva} reports limits directly on the annihilation cross section for the specific channel $\chi\chi\to b\bar b$ as a function of the DM mass. Using the code of Ref.~\cite{Cirelli:2010xx} and following a similar method as that described above for the $\bar p$ spectrum calculation, we verified that the spectrum of continuum gamma rays obtained in our model agrees to within a factor of 2-3 with the gamma ray spectrum resulting from a pure $\chi\chi\to b\bar b$ channel. In what follows we therefore assume that the constraints quoted in~\cite{Ackermann:2013yva} apply to our model directly. In the right panel of Fig.~\ref{fig:dNdE} we plot the differential gamma ray flux from DM annihilation with $m_\chi=6.3$~TeV, $m_\sigma=427$~GeV, and $f=6.2$~TeV reproducing the observed DM relic abundance. The purple line shows the spectrum from the full annihilation process including all dominant partial channels ($\chi\chi\to WW,ZZ,tt,\sigma\sigma,...$), while the blue line shows the spectrum due to $\chi\chi\to bb$ alone. 

We extrapolate the bound to $m_\chi=100$~TeV, using the scaling $m_\chi^{-2}\frac{dN_\gamma}{dE}\sim m_\chi^{-1}$ that applies for photon energies in the FERMI range, $E\lesssim500~{\rm GeV}\ll m_\chi$. The resulting bound is illustrated by the red dashed line in the bottom panel of Fig.~\ref{fig:SEplot}, focusing on a slice in the parameter space with $m_\sigma=3$~TeV.

Stronger, but more model-dependent limits are obtained from ground-based air-Che\-ren\-kov telescopes. The HESS gamma ray observatory reported limits on DM annihilation based on Galactic Center observations~\cite{Abramowski:2011hc}. Due to the background subtraction method of the experiment, the analysis is not sensitive to shallow DM density profiles, and so the results are only applicable under the assumption of a cusp profile such as the Navaro-Frenk-White~\cite{Navarro:1995iw} distribution. Assuming a cusp distribution, neglecting the $\mathcal{O}(1)$ spectral difference between the $\chi\chi\to q\bar q$-induced gamma ray spectrum assumed in~\cite{Abramowski:2011hc} and the actual spectrum in our model, and extrapolating their limits from $m_\chi=10$~TeV up to $m_\chi=100$~TeV, we obtain the bound depicted by the green dashed line in the bottom panel of Fig.~\ref{fig:SEplot}.

Finally, both FERMI~\cite{Ackermann:2012qk} and HESS~\cite{Abramowski:2013ax} reported limits on DM annihilation to a gamma ray line. We calculate the branching fraction $\langle\sigma v\rangle(\chi\chi\to\gamma\gamma)/\langle\sigma v\rangle({\rm total})$ using Eq.~(\ref{eq:xxtogg}). This branching fraction is very small in our model, reminiscent of the result for a heavy Higgs. Consequently the gamma ray line constraint is sub-dominant compared to the continuum emission bounds. We comment that the HESS limit~\cite{Abramowski:2013ax} have recently been used to put significant pressure on supersymmetric Wino dark matter~\cite{Cohen:2013ama, Fan:2013faa}. This situation is not reproduced here; for the Wino example, the strong exclusion is primarily due to the presence of an electromagnetically charged state that is mass-degenerate with the neutral DM particle, amplifying the di-photon annihilation diagram. Without a special construction of this kind, our dilaton-mediated DM scenario passes the line searches unscathed.

%%%%%%%%%%%%%%%%%%%%%%%%%%%%%%%%%
\section{Conclusions}\label{sec:conc}

In this paper we explored the possibility that the dilaton could mediate dark matter annihilation. Such models have the appeal that the couplings are largely determined by scale invariance. The breaking scale of scale invariance $f$ is fixed by requiring that the relic abundance matches the observed value, leaving the dark matter and dilaton masses as the main free parameters. We mapped the relevant $\{f,m_{\chi},m_{\sigma}\}$ parameter space taking the various dark matter annihilation modes into account and imposing unitarity bounds. We showed that large regions of  parameter space, with $f,m_\chi, m_\sigma$ all in the $\sim 1-10$ TeV range, can correctly reproduce the observed relic abundance.  We find an upper bound $f \leq 30-100$ TeV, implying a similar bound on $m_{\sigma ,\chi}$. 

Collider searches for Higgs-like particles, including LHC, Tevatron and LEP analyses, put model dependent lower bounds on $f$ for dilaton masses up to $\sim1$~TeV. The collider bounds exclude dilaton-mediated dark matter for $m_\chi\lesssim200$~GeV. Current direct detection experiments yield similar model dependent exclusions for the lower end of the mass spectrum, requiring $m_{\chi}\gtrsim 300$~GeV for $ m_{\sigma}\lesssim 300$~GeV. The predicted dark matter-nucleon elastic scattering cross section becomes independent of the dark matter mass for heavy dark matter.

Our analysis of indirect detection included antiproton and gamma ray data and shows that the bulk of the parameter space is consistent with the current constraints. A possible signal in high energy cosmic ray antiprotons could appear for favorable cosmic ray propagation scenario for  models with parameters close to a Sommerfeld enhancement resonance. A promising avenue for probing the model all the way to very high DM mass is in high energy ground-based gamma ray measurements, see e.g.~\cite{Funk:2013gxa,Doro:2014sla} for recent reviews. For scalar or vector DM, future gamma ray experiments should detect or exclude the entire parameter space of the model. 

\bigskip
\section*{Acknowledgements} 
We thank Andre Walker-Loud for pointing out the latest lattice QCD hadronic matrix elements.  C.C. thanks the Aspen Center for Physics for its hospitality while part of this work was completed. S.L. thanks the particle theory group at Cornell University, the Institute for Advanced Study in Princeton and the Mainz Institute for Theoretical Physics (MITP) for their hospitality and support while part of this work was completed. K.B. is supported by the DOE grant DE-SC000998. M.C. and C.C. are supported in part by the NSF grant PHY-1316222. SL is supported in part by the National Research Foundation of Korea  grant MEST No. 2012R1A2A2A01045722.

\bigskip
\section*{Note Added} 
While completing this project we became aware of~\cite{Efrati:2014aea} which investigates similar issues. 

\appendix

\section{Collider bounds \label{app:C}}

\begin{figure}[ht]
\begin{center}
\includegraphics[width=.65\textwidth]{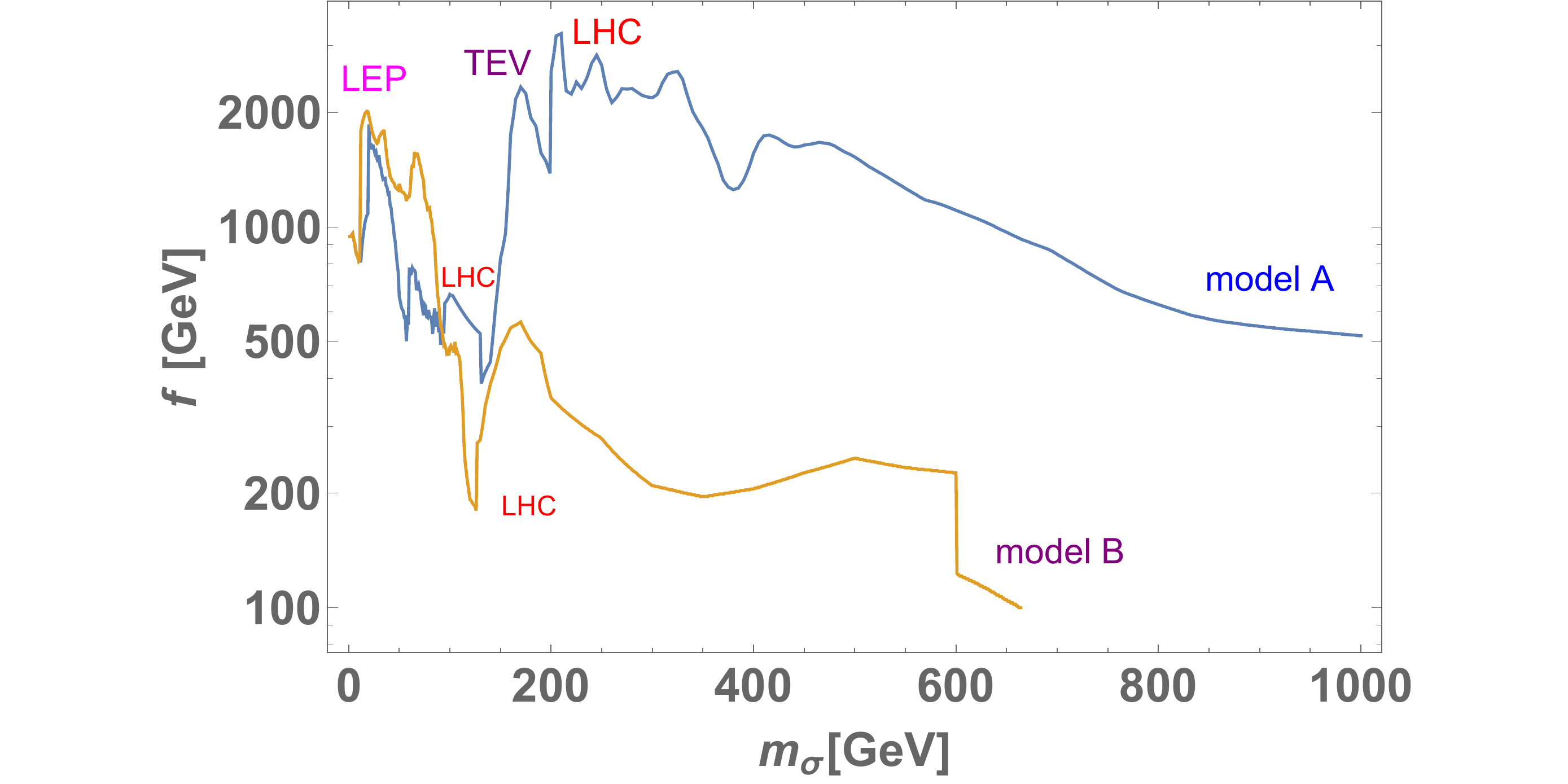}
\end{center}
\caption{95\% C.L. collider exclusion limit on the scale of conformal symmetry breaking, $f$, with respect to $m_\sigma$ for our benchmark models A and B.
}
\label{fig:ColliderBound}
\end{figure} 

As mentioned in the main text, in addition to the direct detection bounds there are also collider bounds from the LHC and earlier experiments. The dilaton (roughly) mimics a Higgs boson, with couplings to massive SM fields suppressed by the factor $v/f$ compared to that of the Higgs and couplings to massless gauge bosons that involve contributions from the matter content of the conformal sector. Collider bounds on the dilaton can thus be obtained by recasting the results of direct production limits from Higgs boson searches. We use the  HiggsBound~\cite{arXiv:0811.4169,arXiv:1102.1898,arXiv:1301.2345}  code version 4.1.2, that incorporates all the currently available experimental analyses from LEP, the Tevatron, and the LHC~\cite{arXiv:0811.4169,arXiv:1102.1898,arXiv:1301.2345}.  

The resulting 
collider bounds on the conformal symmetry breaking scale $f$ as a function of the dilaton mass is presented in Fig.~\ref{fig:ColliderBound} for the two benchmark models A and B defined in Sec.~\ref{sec:mod}. In obtaining these bounds we assumed, for simplicity, no invisible decay channels for the dilaton. We can see that the collider bounds are strongly model dependent: model A has a large coupling to gluons, and thus is very strongly constrained throughout the parameter space relevant for LHC kinematics. Model B has small couplings to gluons and photons, and is only weakly constrained for dilaton masses above 200 GeV. 

The resulting bound on $f$ can be turned into a bound on $m_\chi$ using Fig.~\ref{fig:DMParamSpace}. For example the $f\gtrsim 2$ TeV bound for $m_\sigma \lesssim 400$ GeV in model A implies $m_\chi \gtrsim 300$ GeV, with the exception for a narrow resonance region.

%%%%%%%%%%%%%%%%%%%%%%%%%%%%%%%%%%%%%%
\section{Additional annihilation channels \label{app:A}}

\subsection{Scalar dark matter}

Annihilation to fermions of mass $m_{\psi}$:
\begin{eqnarray}
\sigma v(\chi\chi\rightarrow \bar{\psi}\psi) &\simeq& \frac{m_{\psi}^2m_{\chi}\left(m_{\chi}^2-m_{\psi}^2\right)^{3/2}}{\pi f^4|4m_{\chi}^2-m_{\sigma}^2-i\text{Im}\left(\Pi(4m_{\chi}^2)\right)|^2}
\end{eqnarray}
Annihilation to a real scalar of mass $m_h$:
\begin{eqnarray}
\sigma v(\chi\chi\rightarrow hh) &\simeq& \frac{m_{\chi}m_h^4\sqrt{m_{\chi}^2-m_h^2}}{4\pi f^4|4m_{\chi}^2-m_{\sigma}^2-i\text{Im}\left(\Pi(4m_{\chi}^2)\right)|^2}
\end{eqnarray}
While annihilation to photons is negligible for the relic abundance calculation, it is important for indirect detection.  We get
\begin{eqnarray}\label{eq:xxtogg}
\sigma v(\chi\chi\rightarrow \gamma\gamma) &\simeq&  \frac{3m_{\chi}^6\alpha_{\text{EM}}^2 c_{\text{EM}}^2}{16\pi^3f^4|4m_{\chi}^2-m_{\sigma}^2-i\text{Im}\left(\Pi(4m_{\chi}^2)\right)|^2}
\end{eqnarray}
where $c_{\text{EM}}$ encodes the coupling of photons to dilaton:
\begin{eqnarray}
c_{\text{EM}}&=&\left(F_{W}(x_W)-\sum_f N_c Q_f^2F_f(x_f)+b^{(\text{EM})}_{\text{IR}}-b^{(\text{EM})}_{\text{UV}}\right)\\
x_i&=&\frac{m_i^2}{m_\chi^2}\\
F_W(x)&=&2+3x+3x(2-x)f(x)\\
F_f(x)&=&2x[1+(1-x)f(x)]\\
f(x)&=&\left\{
     \begin{array}{lr}
       \arcsin(1/\sqrt{x})^2 & : x \geq 1 \\
       -\frac{1}{4}\left[\log\left(\frac{1+\sqrt{x-1}}{1-\sqrt{x-1}}\right)-i\pi\right]^2 & : x < 1
     \end{array}
   \right.
\end{eqnarray}

\subsection{Fermion dark matter}

Annihilation to fermions of mass $m_{\psi}$:
\begin{eqnarray}
\sigma v(\bar{\chi}\chi\rightarrow \bar{\psi}\psi) &\simeq& v^2 \frac{m_{\psi}^2m_{\chi}\left(m_{\chi}^2-m_{\psi}^2\right)^{3/2}}{8\pi f^4|4m_{\chi}^2-m_{\sigma}^2-i\text{Im}\left(\Pi(4m_{\chi}^2)\right)|^2}
\end{eqnarray}
Annihilation to a real scalar of mass $m_h$:
\begin{eqnarray}
\sigma v(\bar{\chi}\chi\rightarrow hh) &\simeq& v^2 \frac{m_{\chi}m_h^4\sqrt{m_{\chi}^2-m_h^2}}{32\pi f^4|4m_{\chi}^2-m_{\sigma}^2-i\text{Im}\left(\Pi(4m_{\chi}^2)\right)|^2}
\end{eqnarray}

\subsection{Vector dark matter}

Annihilation to fermions of mass $m_{\psi}$:
\begin{eqnarray}
\sigma v(\chi\chi\rightarrow \bar{\psi}\psi) &\simeq&  \frac{m_{\psi}^2m_{\chi}\left(m_{\chi}^2-m_{\psi}^2\right)^{3/2}}{3\pi f^4|4m_{\chi}^2-m_{\sigma}^2-i\text{Im}\left(\Pi(4m_{\chi}^2)\right)|^2}
\end{eqnarray}
Annihilation to a real scalar of mass $m_h$:
\begin{eqnarray}
\sigma v(\chi\chi\rightarrow hh) &\simeq&  \frac{m_{\chi}m_h^4\sqrt{m_{\chi}^2-m_h^2}}{12\pi f^4|4m_{\chi}^2-m_{\sigma}^2-i\text{Im}\left(\Pi(4m_{\chi}^2)\right)|^2}
\end{eqnarray}

\section{Dilaton decay channels \label{app:B}}

For decay to a real scalar of mass $m_h$ we get
\begin{eqnarray}
\Gamma_{\sigma}(\sigma\rightarrow hh)=\frac{m_h^4}{8\pi m_{\sigma}f^2}\sqrt{1-\frac{4m_h^2}{m_{\sigma}^2}}
\end{eqnarray}
For decay to fermions we get
\begin{eqnarray}
\Gamma_{\sigma}(\sigma \rightarrow \bar{\psi}\psi)&=&\frac{m_{\sigma}m_{\psi}^2}{8\pi f^2}\left(1-\frac{4m_{\psi}^2}{m_{\sigma}^2}\right)^{3/2}
\end{eqnarray}

\bibliographystyle{utphys} 
\bibliography{DarkMatter}

\providecommand{\href}[2]{#2}\begingroup\raggedright\begin{thebibliography}{10}

\bibitem{CGRT}
C.~Csaki, M.~Graesser, L.~Randall, and J.~Terning, ``{Cosmology of brane models
  with radion stabilization},''
  \href{http://dx.doi.org/10.1103/PhysRevD.62.045015}{{\em Phys.Rev.}
  {\bfseries D62} (2000) 045015},
\href{http://arxiv.org/abs/hep-ph/9911406}{{\ttfamily arXiv:hep-ph/9911406
  [hep-ph]}}.
%%CITATION = HEP-PH/9911406;%%.

\bibitem{GRW}
G.~F. Giudice, R.~Rattazzi, and J.~D. Wells, ``{Graviscalars from higher
  dimensional metrics and curvature Higgs mixing},''
  \href{http://dx.doi.org/10.1016/S0550-3213(00)00686-6}{{\em Nucl.Phys.}
  {\bfseries B595} (2001) 250--276},
\href{http://arxiv.org/abs/hep-ph/0002178}{{\ttfamily arXiv:hep-ph/0002178
  [hep-ph]}}.
%%CITATION = HEP-PH/0002178;%%.

\bibitem{CGK}
C.~Csaki, M.~L. Graesser, and G.~D. Kribs, ``{Radion dynamics and electroweak
  physics},'' \href{http://dx.doi.org/10.1103/PhysRevD.63.065002}{{\em
  Phys.Rev.} {\bfseries D63} (2001) 065002},
\href{http://arxiv.org/abs/hep-th/0008151}{{\ttfamily arXiv:hep-th/0008151
  [hep-th]}}.
%%CITATION = HEP-TH/0008151;%%.

\bibitem{CHL}
C.~Csaki, J.~Hubisz, and S.~J. Lee, ``{Radion phenomenology in realistic warped
  space models},'' \href{http://dx.doi.org/10.1103/PhysRevD.76.125015}{{\em
  Phys.Rev.} {\bfseries D76} (2007) 125015},
\href{http://arxiv.org/abs/0705.3844}{{\ttfamily arXiv:0705.3844 [hep-ph]}}.
%%CITATION = ARXIV:0705.3844;%%.

\bibitem{Goldberger:2008zz}
W.~D. Goldberger, B.~Grinstein, and W.~Skiba, ``{Distinguishing the Higgs boson
  from the dilaton at the Large Hadron Collider},''
  \href{http://dx.doi.org/10.1103/PhysRevLett.100.111802}{{\em Phys.Rev.Lett.}
  {\bfseries 100} (2008) 111802},
\href{http://arxiv.org/abs/0708.1463}{{\ttfamily arXiv:0708.1463 [hep-ph]}}.
%%CITATION = ARXIV:0708.1463;%%.

\bibitem{Eshel:2011wz}
Y.~Eshel, S.~J. Lee, G.~Perez, and Y.~Soreq, ``{Shining Flavor and Radion
  Phenomenology in Warped Extra Dimension},''
  \href{http://dx.doi.org/10.1007/JHEP10(2011)015}{{\em JHEP} {\bfseries 1110}
  (2011) 015},
\href{http://arxiv.org/abs/1106.6218}{{\ttfamily arXiv:1106.6218 [hep-ph]}}.
%%CITATION = ARXIV:1106.6218;%%.

\bibitem{Akerib:2013tjd}
{\bfseries LUX Collaboration} Collaboration, D.~Akerib {\em et~al.}, ``{First
  results from the LUX dark matter experiment at the Sanford Underground
  Research Facility},''
  \href{http://dx.doi.org/10.1103/PhysRevLett.112.091303}{{\em Phys.Rev.Lett.}
  {\bfseries 112} (2014) 091303},
\href{http://arxiv.org/abs/1310.8214}{{\ttfamily arXiv:1310.8214
  [astro-ph.CO]}}.
%%CITATION = ARXIV:1310.8214;%%.

\bibitem{Aprile:2012nq}
{\bfseries XENON100 Collaboration} Collaboration, E.~Aprile {\em et~al.},
  ``{Dark Matter Results from 225 Live Days of XENON100 Data},''
  \href{http://dx.doi.org/10.1103/PhysRevLett.109.181301}{{\em Phys.Rev.Lett.}
  {\bfseries 109} (2012) 181301},
\href{http://arxiv.org/abs/1207.5988}{{\ttfamily arXiv:1207.5988
  [astro-ph.CO]}}.
%%CITATION = ARXIV:1207.5988;%%.

\bibitem{Agnese:2013rvf}
{\bfseries CDMS Collaboration} Collaboration, R.~Agnese {\em et~al.},
  ``{Silicon Detector Dark Matter Results from the Final Exposure of CDMS
  II},'' \href{http://dx.doi.org/10.1103/PhysRevLett.111.251301}{{\em
  Phys.Rev.Lett.} {\bfseries 111} (2013) 251301},
\href{http://arxiv.org/abs/1304.4279}{{\ttfamily arXiv:1304.4279 [hep-ex]}}.
%%CITATION = ARXIV:1304.4279;%%.

\bibitem{Bai:2009ms}
Y.~Bai, M.~Carena, and J.~Lykken, ``{Dilaton-assisted Dark Matter},''
  \href{http://dx.doi.org/10.1103/PhysRevLett.103.261803}{{\em Phys.Rev.Lett.}
  {\bfseries 103} (2009) 261803},
\href{http://arxiv.org/abs/0909.1319}{{\ttfamily arXiv:0909.1319 [hep-ph]}}.
%%CITATION = ARXIV:0909.1319;%%.

\bibitem{Agashe:2009ja}
K.~Agashe, K.~Blum, S.~J. Lee, and G.~Perez, ``{Astrophysical Implications of a
  Visible Dark Matter Sector from a Custodially Warped-GUT},''
  \href{http://dx.doi.org/10.1103/PhysRevD.81.075012}{{\em Phys.Rev.}
  {\bfseries D81} (2010) 075012},
\href{http://arxiv.org/abs/0912.3070}{{\ttfamily arXiv:0912.3070 [hep-ph]}}.
%%CITATION = ARXIV:0912.3070;%%.

\bibitem{Bellazzini:2012vz}
B.~Bellazzini, C.~Csaki, J.~Hubisz, J.~Serra, and J.~Terning, ``{A Higgslike
  Dilaton},'' \href{http://dx.doi.org/10.1140/epjc/s10052-013-2333-x}{{\em
  Eur.Phys.J.} {\bfseries C73} (2013) 2333},
\href{http://arxiv.org/abs/1209.3299}{{\ttfamily arXiv:1209.3299 [hep-ph]}}.
%%CITATION = ARXIV:1209.3299;%%.

\bibitem{Chacko:2012sy}
Z.~Chacko and R.~K. Mishra, ``{Effective Theory of a Light Dilaton},''
  \href{http://dx.doi.org/10.1103/PhysRevD.87.115006}{{\em Phys.Rev.}
  {\bfseries D87} no.~11, (2013) 115006},
\href{http://arxiv.org/abs/1209.3022}{{\ttfamily arXiv:1209.3022 [hep-ph]}}.
%%CITATION = ARXIV:1209.3022;%%.

\bibitem{Chacko}
Z.~Chacko, R.~K. Mishra, and D.~Stolarski, ``{Dynamics of a Stabilized Radion
  and Duality},'' \href{http://dx.doi.org/10.1007/JHEP09(2013)121}{{\em JHEP}
  {\bfseries 1309} (2013) 121},
\href{http://arxiv.org/abs/1304.1795}{{\ttfamily arXiv:1304.1795 [hep-ph]}}.
%%CITATION = ARXIV:1304.1795;%%.

\bibitem{BCHST}
B.~Bellazzini, C.~Csaki, J.~Hubisz, J.~Serra, and J.~Terning, ``{A Naturally
  Light Dilaton and a Small Cosmological Constant},''
  \href{http://dx.doi.org/10.1140/epjc/s10052-014-2790-x}{{\em Eur.Phys.J.}
  {\bfseries C74} (2014) 2790},
\href{http://arxiv.org/abs/1305.3919}{{\ttfamily arXiv:1305.3919 [hep-th]}}.
%%CITATION = ARXIV:1305.3919;%%.

\bibitem{Rattazzi}
F.~Coradeschi, P.~Lodone, D.~Pappadopulo, R.~Rattazzi, and L.~Vitale, ``{A
  naturally light dilaton},''
  \href{http://dx.doi.org/10.1007/JHEP11(2013)057}{{\em JHEP} {\bfseries 1311}
  (2013) 057},
\href{http://arxiv.org/abs/1306.4601}{{\ttfamily arXiv:1306.4601 [hep-th]}}.
%%CITATION = ARXIV:1306.4601;%%.

\bibitem{Ade:2013zuv}
{\bfseries Planck Collaboration} Collaboration, P.~Ade {\em et~al.}, ``{Planck
  2013 results. XVI. Cosmological parameters},''
  \href{http://dx.doi.org/10.1051/0004-6361/201321591}{{\em Astron.Astrophys.}
  (2014) },
\href{http://arxiv.org/abs/1303.5076}{{\ttfamily arXiv:1303.5076
  [astro-ph.CO]}}.
%%CITATION = ARXIV:1303.5076;%%.

\bibitem{griest1990unitarity}
K.~Griest and M.~Kamionkowski, ``Unitarity limits on the mass and radius of
  dark-matter particles,'' {\em Phys. Rev. Lett.} {\bfseries 64} no.~6, (1990)
  615.

\bibitem{Feng:2010zp}
J.~L. Feng, M.~Kaplinghat, and H.-B. Yu, ``{Sommerfeld Enhancements for Thermal
  Relic Dark Matter},''
  \href{http://dx.doi.org/10.1103/PhysRevD.82.083525}{{\em Phys.Rev.}
  {\bfseries D82} (2010) 083525},
\href{http://arxiv.org/abs/1005.4678}{{\ttfamily arXiv:1005.4678 [hep-ph]}}.
%%CITATION = ARXIV:1005.4678;%%.

\bibitem{Fox:2011pm}
P.~J. Fox, R.~Harnik, J.~Kopp, and Y.~Tsai, ``{Missing Energy Signatures of
  Dark Matter at the LHC},''
  \href{http://dx.doi.org/10.1103/PhysRevD.85.056011}{{\em Phys.Rev.}
  {\bfseries D85} (2012) 056011},
\href{http://arxiv.org/abs/1109.4398}{{\ttfamily arXiv:1109.4398 [hep-ph]}}.
%%CITATION = ARXIV:1109.4398;%%.

\bibitem{Junnarkar:2013ac}
P.~Junnarkar and A.~Walker-Loud, ``{Scalar strange content of the nucleon from
  lattice QCD},'' \href{http://dx.doi.org/10.1103/PhysRevD.87.114510}{{\em
  Phys.Rev.} {\bfseries D87} no.~11, (2013) 114510},
\href{http://arxiv.org/abs/1301.1114}{{\ttfamily arXiv:1301.1114 [hep-lat]}}.
%%CITATION = ARXIV:1301.1114;%%.

\bibitem{Bai:2010hh}
Y.~Bai, P.~J. Fox, and R.~Harnik, ``{The Tevatron at the Frontier of Dark
  Matter Direct Detection},''
  \href{http://dx.doi.org/10.1007/JHEP12(2010)048}{{\em JHEP} {\bfseries 1012}
  (2010) 048},
\href{http://arxiv.org/abs/1005.3797}{{\ttfamily arXiv:1005.3797 [hep-ph]}}.
%%CITATION = ARXIV:1005.3797;%%.

\bibitem{Ellis:2008hf}
J.~R. Ellis, K.~A. Olive, and C.~Savage, ``{Hadronic Uncertainties in the
  Elastic Scattering of Supersymmetric Dark Matter},''
  \href{http://dx.doi.org/10.1103/PhysRevD.77.065026}{{\em Phys.Rev.}
  {\bfseries D77} (2008) 065026},
\href{http://arxiv.org/abs/0801.3656}{{\ttfamily arXiv:0801.3656 [hep-ph]}}.
%%CITATION = ARXIV:0801.3656;%%.

\bibitem{Blum:2010nx}
K.~Blum, ``{Cosmic ray propagation time scales: lessons from radioactive nuclei
  and positron data},''
  \href{http://dx.doi.org/10.1088/1475-7516/2011/11/037}{{\em JCAP} {\bfseries
  1111} (2011) 037},
\href{http://arxiv.org/abs/1010.2836}{{\ttfamily arXiv:1010.2836
  [astro-ph.HE]}}.
%%CITATION = ARXIV:1010.2836;%%.

\bibitem{Hisano:2003ec}
J.~Hisano, S.~Matsumoto, and M.~M. Nojiri, ``{Explosive dark matter
  annihilation},'' \href{http://dx.doi.org/10.1103/PhysRevLett.92.031303}{{\em
  Phys.Rev.Lett.} {\bfseries 92} (2004) 031303},
\href{http://arxiv.org/abs/hep-ph/0307216}{{\ttfamily arXiv:hep-ph/0307216
  [hep-ph]}}.
%%CITATION = HEP-PH/0307216;%%.

\bibitem{Hisano:2004ds}
J.~Hisano, S.~Matsumoto, M.~M. Nojiri, and O.~Saito, ``{Non-perturbative effect
  on dark matter annihilation and gamma ray signature from galactic center},''
  \href{http://dx.doi.org/10.1103/PhysRevD.71.063528}{{\em Phys.Rev.}
  {\bfseries D71} (2005) 063528},
\href{http://arxiv.org/abs/hep-ph/0412403}{{\ttfamily arXiv:hep-ph/0412403
  [hep-ph]}}.
%%CITATION = HEP-PH/0412403;%%.

\bibitem{Cassel:2009wt}
S.~Cassel, ``{Sommerfeld factor for arbitrary partial wave processes},''
  \href{http://dx.doi.org/10.1088/0954-3899/37/10/105009}{{\em J.Phys.}
  {\bfseries G37} (2010) 105009},
\href{http://arxiv.org/abs/0903.5307}{{\ttfamily arXiv:0903.5307 [hep-ph]}}.
%%CITATION = ARXIV:0903.5307;%%.

\bibitem{Slatyer:2009vg}
T.~R. Slatyer, ``{The Sommerfeld enhancement for dark matter with an excited
  state},'' \href{http://dx.doi.org/10.1088/1475-7516/2010/02/028}{{\em JCAP}
  {\bfseries 1002} (2010) 028},
\href{http://arxiv.org/abs/0910.5713}{{\ttfamily arXiv:0910.5713 [hep-ph]}}.
%%CITATION = ARXIV:0910.5713;%%.

\bibitem{Mayorov:2013coa}
A.~Mayorov, O.~Adriani, G.~Barbarino, G.~Bazilevskaia, R.~Belotti, {\em
  et~al.}, ``{Antiprotons of galactic cosmic radiation in the PAMELA
  experiment},''
\href{http://dx.doi.org/10.3103/S1062873813050389}{{\em
  Bull.Russ.Acad.Sci.Phys.} {\bfseries 77} (2013) 602--605}.
%%CITATION = BUPSA,77,602;%%.

\bibitem{Katz:2009yd}
B.~Katz, K.~Blum, and E.~Waxman, ``{What can we really learn from positron flux
  'anomalies'?},'' {\em Mon.Not.Roy.Astron.Soc.} {\bfseries 405} (2010) 1458,
\href{http://arxiv.org/abs/0907.1686}{{\ttfamily arXiv:0907.1686
  [astro-ph.HE]}}.
%%CITATION = ARXIV:0907.1686;%%.

\bibitem{Cirelli:2010xx}
M.~Cirelli, G.~Corcella, A.~Hektor, G.~Hutsi, M.~Kadastik, {\em et~al.},
  ``{PPPC 4 DM ID: A Poor Particle Physicist Cookbook for Dark Matter Indirect
  Detection},'' \href{http://dx.doi.org/10.1088/1475-7516/2012/10/E01,
  10.1088/1475-7516/2011/03/051}{{\em JCAP} {\bfseries 1103} (2011) 051},
\href{http://arxiv.org/abs/1012.4515}{{\ttfamily arXiv:1012.4515 [hep-ph]}}.
%%CITATION = ARXIV:1012.4515;%%.

\bibitem{Kounine:2012ega}
A.~Kounine, ``{The Alpha Magnetic Spectrometer on the International Space
  Station},''
\href{http://dx.doi.org/10.1142/S0218301312300056}{{\em Int.J.Mod.Phys.}
  {\bfseries E21} no.~08, (2012) 1230005}.
%%CITATION = IMPAE,E21,1230005;%%.

\bibitem{Ginzburg:1990sk}
V.~Ginzburg, V.~Dogiel, V.~Berezinsky, S.~Bulanov, and V.~Ptuskin,
``{Astrophysics of cosmic rays, Amsterdam, Netherlands: North-Holland (1990)
  534 p},''.
%%CITATION = INSPIRE-307564;%%.

\bibitem{Ackermann:2013yva}
{\bfseries Fermi-LAT Collaboration} Collaboration, M.~Ackermann {\em et~al.},
  ``{Dark matter constraints from observations of 25 Milky Way satellite
  galaxies with the Fermi Large Area Telescope},''
  \href{http://dx.doi.org/10.1103/PhysRevD.89.042001}{{\em Phys.Rev.}
  {\bfseries D89} no.~4, (2014) 042001},
\href{http://arxiv.org/abs/1310.0828}{{\ttfamily arXiv:1310.0828
  [astro-ph.HE]}}.
%%CITATION = ARXIV:1310.0828;%%.

\bibitem{Abramowski:2011hc}
{\bfseries H.E.S.S.Collaboration} Collaboration, A.~Abramowski {\em et~al.},
  ``{Search for a Dark Matter annihilation signal from the Galactic Center halo
  with H.E.S.S},'' \href{http://dx.doi.org/10.1103/PhysRevLett.106.161301}{{\em
  Phys.Rev.Lett.} {\bfseries 106} (2011) 161301},
\href{http://arxiv.org/abs/1103.3266}{{\ttfamily arXiv:1103.3266
  [astro-ph.HE]}}.
%%CITATION = ARXIV:1103.3266;%%.

\bibitem{Navarro:1995iw}
J.~F. Navarro, C.~S. Frenk, and S.~D. White, ``{The Structure of cold dark
  matter halos},'' \href{http://dx.doi.org/10.1086/177173}{{\em Astrophys.J.}
  {\bfseries 462} (1996) 563--575},
\href{http://arxiv.org/abs/astro-ph/9508025}{{\ttfamily arXiv:astro-ph/9508025
  [astro-ph]}}.
%%CITATION = ASTRO-PH/9508025;%%.

\bibitem{Ackermann:2012qk}
{\bfseries LAT Collaboration} Collaboration, M.~Ackermann {\em et~al.},
  ``{Fermi LAT Search for Dark Matter in Gamma-ray Lines and the Inclusive
  Photon Spectrum},'' \href{http://dx.doi.org/10.1103/PhysRevD.86.022002}{{\em
  Phys.Rev.} {\bfseries D86} (2012) 022002},
\href{http://arxiv.org/abs/1205.2739}{{\ttfamily arXiv:1205.2739
  [astro-ph.HE]}}.
%%CITATION = ARXIV:1205.2739;%%.

\bibitem{Abramowski:2013ax}
{\bfseries H.E.S.S. Collaboration} Collaboration, A.~Abramowski {\em et~al.},
  ``{Search for photon line-like signatures from Dark Matter annihilations with
  H.E.S.S},'' \href{http://dx.doi.org/10.1103/PhysRevLett.110.041301}{{\em
  Phys.Rev.Lett.} {\bfseries 110} (2013) 041301},
\href{http://arxiv.org/abs/1301.1173}{{\ttfamily arXiv:1301.1173
  [astro-ph.HE]}}.
%%CITATION = ARXIV:1301.1173;%%.

\bibitem{Cohen:2013ama}
T.~Cohen, M.~Lisanti, A.~Pierce, and T.~R. Slatyer, ``{Wino Dark Matter Under
  Siege},'' \href{http://dx.doi.org/10.1088/1475-7516/2013/10/061}{{\em JCAP}
  {\bfseries 1310} (2013) 061},
\href{http://arxiv.org/abs/1307.4082}{{\ttfamily arXiv:1307.4082}}.
%%CITATION = ARXIV:1307.4082;%%.

\bibitem{Fan:2013faa}
J.~Fan and M.~Reece, ``{In Wino Veritas? Indirect Searches Shed Light on
  Neutralino Dark Matter},''
  \href{http://dx.doi.org/10.1007/JHEP10(2013)124}{{\em JHEP} {\bfseries 1310}
  (2013) 124},
\href{http://arxiv.org/abs/1307.4400}{{\ttfamily arXiv:1307.4400 [hep-ph]}}.
%%CITATION = ARXIV:1307.4400;%%.

\bibitem{Funk:2013gxa}
S.~Funk, ``{Indirect Detection of Dark Matter with gamma rays},''
\href{http://arxiv.org/abs/1310.2695}{{\ttfamily arXiv:1310.2695
  [astro-ph.HE]}}.
%%CITATION = ARXIV:1310.2695;%%.

\bibitem{Doro:2014sla}
M.~Doro, ``{A decade of dark matter searches with ground-based Cherenkov
  telescopes},'' \href{http://dx.doi.org/10.1016/j.nima.2013.12.010}{{\em
  Nucl.Instrum.Meth.} {\bfseries A742} (2014) 99--106},
\href{http://arxiv.org/abs/1404.5017}{{\ttfamily arXiv:1404.5017
  [astro-ph.HE]}}.
%%CITATION = ARXIV:1404.5017;%%.

\bibitem{Efrati:2014aea}
A.~Efrati, E.~Kuflik, S.~Nussinov, Y.~Soreq, and T.~Volansky, ``{Constraining
  the Higgs-Dilaton with LHC and Dark Matter Searches},''
\href{http://arxiv.org/abs/1410.2225}{{\ttfamily arXiv:1410.2225 [hep-ph]}}.
%%CITATION = ARXIV:1410.2225;%%.

\bibitem{arXiv:0811.4169}
P.~Bechtle, O.~Brein, S.~Heinemeyer, G.~Weiglein, and K.~E. Williams,
  ``{HiggsBounds: Confronting Arbitrary Higgs Sectors with Exclusion Bounds
  from LEP and the Tevatron},''
  \href{http://dx.doi.org/10.1016/j.cpc.2009.09.003}{{\em Comput. Phys.
  Commun.} {\bfseries 181} (2010) 138--167},
\href{http://arxiv.org/abs/0811.4169}{{\ttfamily arXiv:0811.4169 [hep-ph]}}.
%%CITATION = 0811.4169;%%.

\bibitem{arXiv:1102.1898}
P.~Bechtle, O.~Brein, S.~Heinemeyer, G.~Weiglein, and K.~E. Williams,
  ``{HiggsBounds 2.0.0: Confronting Neutral and Charged Higgs Sector
  Predictions with Exclusion Bounds from LEP and the Tevatron},''
  \href{http://dx.doi.org/10.1016/j.cpc.2011.07.015}{{\em Comput. Phys.
  Commun.} {\bfseries 182} (2011) 2605--2631},
\href{http://arxiv.org/abs/1102.1898}{{\ttfamily arXiv:1102.1898 [hep-ph]}}.
%%CITATION = 1102.1898;%%.

\bibitem{arXiv:1301.2345}
P.~Bechtle {\em et~al.}, ``{Recent Developments in HiggsBounds and a Preview of
  HiggsSignals},'' {\em PoS} {\bfseries CHARGED2012} (2012) 024,
\href{http://arxiv.org/abs/1301.2345}{{\ttfamily arXiv:1301.2345 [hep-ph]}}.
%%CITATION = 1301.2345;%%.

\end{thebibliography}\endgroup

%\end{thebibliography}

\end{document}